\documentclass[hidelinks]{article}
\usepackage{amsmath}
\usepackage{amsfonts}
\usepackage{comment}
\usepackage{hyperref}
\usepackage[square,sort&compress,numbers]{natbib}
\usepackage{empheq}
\usepackage[scr=euler]{mathalfa}

\newcommand*\widefbox[1]{\fbox{\hspace{2em}#1\hspace{2em}}}

\def\p{\partial}
\def\O{\mathcal{O}}
\def\a{\alpha}
\def\b{\beta}
\def\g{\gamma}
\def\r{\rightarrow}

\def\l{\lambda}
\def\d{\delta}
\def\s{\sigma}
\def\e{\epsilon}
\def\om{\omega}
\def\L{\mathcal{L}}
\def\lt{\tilde{\lambda}}
\def\mut{\tilde{\mu}}

\newcommand{\be}{\begin{equation}}
\newcommand{\ee}{\end{equation}}
\newcommand{\bea}{\begin{eqnarray}}
\newcommand{\eea}{\end{eqnarray}}
\newcommand{\bi}{\begin{itemize}}
\newcommand{\ei}{\end{itemize}}

\makeatletter \@addtoreset{equation}{section} \makeatother

\title{\bf  The holographic interpretation of $J \bar T$-deformed CFTs
\vspace{4mm}
}

\author{Adam Bzowski$^{\flat}$ and
Monica Guica$^{\flat\S\dag}$ \\
\\\vspace{2mm}
${}^\flat$\emph{\normalsize Institut de Physique Th\'eorique, CEA Saclay, CNRS},
\emph{91191 Gif-sur-Yvette, France}\\\vspace{2mm}
${}^\S$\emph{\normalsize Department of Physics and Astronomy, Uppsala University,   SE-751 08 Uppsala, Sweden}\\\vspace{2mm}
${}^\dag$\emph{\normalsize Nordita, Stockholm University and KTH Royal Institute of Technology,}\\\vspace{2mm}
\emph{\normalsize Roslagstullsbacken 23, SE-106 91 Stockholm, Sweden}
}

\date{}

\begin{document}

\maketitle

\begin{abstract}
\vskip3mm

Recently, a non-local yet possibly UV-complete quantum field theory has been constructed by deforming a two-dimensional CFT by the composite operator $J \bar T$, where $J$ is a chiral $U(1)$ current and $\bar T$ is a component of the stress tensor.  Assuming the original CFT was a holographic CFT, we work out the holographic dual of its $J \bar T$ deformation.  We find that the dual spacetime is still AdS$_3$,  but with modified boundary conditions that mix the metric and the  Chern-Simons gauge field dual to the $U(1)$ current. We show that when the coefficient of the chiral anomaly  for $J$ vanishes, the energy and thermodynamics of black holes obeying these modified boundary conditions precisely reproduce the previously derived field theory spectrum and thermodynamics. Our proposed holographic dictionary can also reproduce the field-theoretical spectrum  in presence of the chiral anomaly, upon a certain assumption that we justify. 
%
%
%
 The  asymptotic symmetry group associated to these boundary conditions consists of two copies of the Virasoro  and one copy of the $U(1)$ Ka\v{c}-Moody algebra, just as before the deformation; the only effect of the latter is to modify the spacetime dependence of the right-moving Virasoro generators, whose action  becomes state-dependent and effectively non-local. 
\end{abstract}

\tableofcontents

\section{Introduction}

Recently, an interesting class  of  irrelevant deformations of  two-dimensional quantum field theories has been uncovered \cite{Smirnov:2016lqw}, where the deforming operator is a certain bilinear combination of  conserved currents. Thanks to this special form, the resulting deformed QFT is, in a sense, solvable  - for example, it is possible to 
determine the finite-size spectrum and thermodynamics at finite deformation parameter in terms of the original QFT data. 

The best studied of these deformations is the so-called $T\bar T$ deformation of two-dimensional CFTs, for which the  spectrum takes a universal form \cite{Smirnov:2016lqw,Cavaglia:2016oda}. The $T \bar T$ deformation can be recast in S-matrix language \cite{Dubovsky:2017cnj}, and the  form of the resulting S-matrix suggests that the deformed theory is UV complete, even though it does not possess a usual, local UV fixed point. Theories with such high-energy behaviour have been termed ``asymptotically fragile'', and they open up a whole new set of possible ultraviolet behaviours of QFTs  \cite{Dubovsky:2013ira}. One example is provided by the worldsheet theory of the bosonic string, which has been studied from this perspective in  \cite{Dubovsky:2012wk, Dubovsky:2012sh, Caselle:2013dra}. The $T \bar T$ deformation has also found an interesting application in holography, as the field theory dual to a finite bulk cutoff for the metric fluctuations \cite{McGough:2016lol} (see also \cite{Shyam:2017znq,Kraus:2018xrn,Cottrell:2018skz}) and, in a somewhat modified form, as holographic dual to a linear dilaton background \cite{Giveon:2017nie,Giveon:2017myj,Asrat:2017tzd,Giribet:2017imm}. The partition function of $T\bar T$ deformed QFTs on various domains has been recently studied in \cite{Cardy:2018sdv}, and  correlation functions in a certain large central charge limit have been computed in \cite{Aharony:2018vux}. 

Another potentially interesting  deformation in the Smirnov-Zamolodchikov class  corresponds to  a two-dimensional CFT deformed by an irrelevant double-trace operator of the schematic form  \cite{Guica:2017lia} 
\be
S= S_{CFT} + \mu \int  d^2 z \,J \bar T \label{jtbardef}
\ee
where $\bar T$ represents the stress tensor and $J$ is a $U(1)$ current. This deformation preserves an $SL(2,\mathbb{R})\times U(1)$ subgroup of the orginal conformal group, and it may be relevant for the holographic understanding of extremal black holes \cite{Guica:2008mu,ElShowk:2011cm}. For $J$  purely chiral or antichiral, the deformed spectrum again takes a universal form; the deformation is non-trivial for chiral $J$,  while for $J$ antichiral no modification away from the CFT spectrum was observed. 

It should be noted that the original analysis of the spectrum of $J\bar T$-deformed CFTs performed in \cite{Guica:2017lia} holds in the limit of vanishing chiral anomaly for the current $J$. The authors of \cite{Chakraborty:2018vja}, who studied a single-trace analogue of the $J\bar T$ deformation, give an argument for what the spectrum of  $J\bar T$-deformed CFTs should be in presence of the chiral anomaly. This spectrum has a number of interesting new features, such as an upper bound on the allowed energies, similar to the behaviour of the $T\bar T$-deformed spectrum at negative $\mu$.

In this article, we  model the $J\bar T$ deformation of two-dimensional CFTs in holography. The minimal ingredients in the bulk are three-dimensional Einstein gravity, which provides a holographic dual for the stress tensor, coupled to a  Chern-Simons gauge-field, dual 
to the $U(1)$ current. By choosing the sign of the Chern-Simons coupling, the current can be made chiral or anti-chiral. We will exclusively concentrate on the non-trivial chiral $J$ case. 

Since  there are no dynamical degrees of freedom in this system, all the bulk solutions are locally AdS$_3$. The only effect of the double-trace deformation is to change the asymptotic boundary conditions imposed on the bulk fields, from Dirichlet to mixed ones. The simplest way to derive the new holographic sources and expectation values is by analysing the variational principle in presence of the deformation. We find that  the asymptotic boundary conditions on the metric are very similar to the ``new boundary conditions for AdS$_3$'' proposed in \cite{Compere:2013bya}; however, since  our setup contains an additional gauge field, the allowed excitations are no longer restricted to be chiral. We also find that the expectation values in the deformed theory are related in a simple way to the expectation values in the original CFT in presence of non-trivial sources, which can be computed using the ``usual'' AdS$_3$/CFT$_2$ holographic dictionary (i.e., with Dirichlet boundary conditions on the metric). We check this dictionary by showing that the energy 
 of black holes with these boundary conditions  reproduce the spectrum 
 previously derived from purely field-theoretical considerations.

The only subtle point in our analysis concerns properly taking into account  the effect of the chiral anomaly
on the expectation value of the $U(1)$ current  along the flow. This is a field-theoretical question whose answer is currently not well understood. Our analysis is thus split into two cases: i) that of vanishing chiral anomaly, where the holographic dictionary is perfectly well understood and it matches both the field-theory spectrum and thermodynamics in this limit\footnote{While one may object that in this case the CFT we are deforming is either trivial or non-unitary, our point of view is that unitarity constraints do not play an essential role in the $N \r \infty$ limit in which we are working and are thus not expected to significantly alter the dictionary we derive. Our exact match of the holographic to the field-theoretical spectrum for $k=0$ should thus be viewed as an interesting limit, for which the flow is under full control, of the more physical $k \neq 0$ case.}, and ii) that of non-vanishing anomaly, where the behaviour of the current is under less control, which in turn impacts  our understanding of the holographic dictionary. By drawing an analogy with the case of a $J\bar J$-type deformation that we can work out exactly (see appendix \ref{cc}), in section \ref{jtrev} we arrive at an intuitive picture for the role of the anomaly in the flow. If this picture is correct, then it suggests a particular identification in the holographic dictionary for non-zero anomaly, which in turn perfectly reproduces the field-theoretical prediction for the spectrum.


Note that even though the above discussion takes place in the context of holography,  our calculations can also be viewed from a purely field-theoretical point of view, as a way to compute  expectation values in the deformed theory at large $N$ in terms of expectation values in the original CFT. While here we concentrate on the one-point functions of the stress tensor and the current, in principle our method can be used to compute arbitrary correlators in the deformed theory in terms of CFT ones.

As mentioned above, the $J\bar T$ deformation breaks the two-dimensional conformal group $SL(2,\mathbb{R})_L \times SL(2,\mathbb{R})_R$  down to an $SL(2,\mathbb{R})_L \times U(1)_R$ subgroup; additionally, there is the chiral $U(1)_J$ symmetry generated by the current. An interesting question is whether these global symmetries are enhanced to infinite-dimensional ones, as is common in two dimensions. There are two ways to address this question: either by constructing the infinite set of conserved charges explicitly using the special properties of the stress tensor and the current, as in \cite{Hofman:2011zj}, or by studying the asymptotic symmetries of the dual spacetime, as in  \cite{Brown:1986nw}. In the original CFT, either method can be used to show that the $SL(2,\mathbb{R})_L$ and $U(1)_J$ symmetries are enhanced to a left-moving Virasoro - Ka\v{c}-Moody algebra, while the $SL(2,\mathbb{R})_R$ is enhanced to a right-moving Virasoro algebra. 
In the deformed CFT, we use both methods to show that there is a similar infinite-dimensional enhancement of the  global symmetries to a Virasoro $\times$ Virasoro $\times$ $U(1)$ Ka\v{c}-Moody algebra; 
 the only  change is that the argument of the  right-moving Virasoro generators is shifted by a state-dependent function of the left-moving  boundary coordinate. 


The plan of the paper is as follows. We start section \ref{dtrdef} by reviewing the effect of double-trace deformations in holography from a path integral approach, which is equivalent to the variational approach. We also review the basics of the $J\bar{T}$ deformation, including the effects of the chiral anomaly studied in \cite{Chakraborty:2018vja}. We then apply the variational approach to the specific case of the $J \bar T$ deformation and obtain the deformed  sources and expectation values in terms of the original ones. In section \ref{holodict}, after a quick review of the usual AdS$_3$/CFT$_2$ dictionary,  we find the asymptotic expansion of the bulk fields that corresponds to the new boundary conditions and propose an expression for the holographic expectation values, separately for the case of zero and non-zero chiral anomaly. In section \ref{checkp}, we check this holographic dictionary by
showing that the thermodynamics and  conserved charges of black holes obeying the new asymptotics agree with the field theoretical results. 
Next, we use  holography to construct an infinite set of conserved charges and compute their 
 associated asymptotic symmetry algebra. We end with a discussion and future directions. 
In the appendix, we study a simple example of a $J\bar J$ flow from both the field-theoretical and the holographic perspective, concentrating on the role of the chiral anomaly.

\section{Effect of the double-trace deformation \label{dtrdef}}

The effect of multitrace deformations in the context of the AdS/CFT correspondence  \cite{Witten:2001ua}  has been extensively studied. 
At large $N$, as far as the low-lying single-trace operators\footnote{The effect of the deformation beyond the leading order in $1/N$ and for operators other than single-trace ones has been recently studied in \cite{Giombi:2018vtc}.} are concerned, such deformations simply correspond to changing the asymptotic boundary conditions on their dual supergravity field. 
At the level of the classical supergravity action, the new boundary conditions can be easily read off  by studying the variational principle in presence of the deformation. 

Most of the literature on the subject is concerned with deformations constructed from scalar operators of dimension smaller 
than $d/2$, such that the resulting multitrace operator is relevant or marginal. This ensures that the ultraviolet regime of the theory is under control; in the dual picture, the deformation of AdS is normalizable (though only visible at $1/N$ order), and so also under control.  Note that the bulk field dual to such an operator  in the original CFT  will be quantized with Neumann, also known as alternate, boundary conditions. 

The $J \bar T$ deformation differs in several respects from the usual case. First, the deformation is irrelevant; we will nevertheless consider it, because the resulting theory is expected to be UV complete.  On the bulk side, we  deform the gravitational theory in the usual quantization (i.e., with Dirichlet  boundary conditions for the metric). The resulting mixed boundary conditions  involve fluctuations of the non-normalizable mode of the metric and of the Chern-Simons  gauge field; however, since none of these modes is dynamical, the asymptotic  geometry is still locally AdS$_3$. This lack of backreaction of the deformation on the local geometry  is likely related to the UV-completeness of the dual theory. 

In this section, we start by reviewing the path integral derivation of the change in boundary conditions induced by the double-trace deformation, following \cite{Gubser:2002vv}, and how the same result is recovered in the variational approach. We then apply the variational approach to the $J \bar T$ deformed-CFT  and read off the new  sources and expectation values in terms of the old ones, which in principle gives us the full large $N$ holographic dictionary for this theory.

\subsection{Review of double-trace deformations in holography \label{revdth}}

Let us review how the holographic data change under general double-trace deformations of a holographic (large $N$) CFT, following  \cite{Gubser:2002vv}. In the case usually considered,  a term of the form

\be
 S_{d.tr} = \int d^d x \, \mathcal{L}_{d.tr} (\mathcal{O_A})
\ee
is added to the CFT action, 
where the $\O_A$ are scalar operators dual to supergravity fields, whose  correlation functions factorize at large $N$. 
We will specifically be interested  in the case in which $\mathcal{L}_{d.tr}$ is a bilinear in the operators of interest
\be
\mathcal{L}_{d.tr.} = \frac{1}{2} \, \mu^{AB} \O_A \O_B \label{dtrl}
\ee
where $\mu^{AB}$ is  a constant matrix. The generating functional in the deformed theory is 

\be
e^{- W_\mu [\,\tilde{\mathscr{I}}^A]} = \int \mathcal{D} \varphi \, e^{-S[\varphi] + \int\tilde{\mathscr{I}}^A \O_A - \frac{1}{2} \int \mu^{AB} \O_A \O_B}
\ee
where $\tilde{\mathscr{I}}^A$ denote the sources in the deformed theory that couple to $\O_A$ and $\varphi$ denotes the fundamental degrees of freedom in the CFT, over which the path integral is performed, weighted by the action $S[\varphi]$.   Using the identity
\be
1 = \sqrt{\det{\mu^{-1}}} \int \mathcal{D}\tilde\s^A e^{\frac{1}{2} \int \tilde\s^A (\mu^{-1})_{AB} \tilde\s^B}
\ee
shifting the integration variable as $\tilde \s^A = \s^A - \tilde{\mathscr{I}}^A + \mu^{AB} \O_B$ and using large $N$ factorization, one finds

\be
e^{- W_\mu [\,\tilde{\mathscr{I}}^A]} = \int \mathcal{D} \s^A \, e^{-W[ \s^A]  + \frac{1}{2} \int( \s^A - \tilde{\mathscr{I}}^A) (\mu^{-1})_{AB} ( \s^B - \tilde{\mathscr{I}}^B) }
\ee
Note that $\s^A \equiv \,\mathscr{I}^A$ plays the role of source in the undeformed theory. 
Evaluating the above integral via saddle point, one finds the latter occurs at

\be
\left. - \frac{\d W[ \s^A]}{\d  \s^A} \right|_{ \s^A_*}=  \langle \O_A \rangle = -  (\mu^{-1})_{AB} ( \s_*^B - \tilde{\mathscr{I}}^B)
\ee
and 
\be
 \s^A_* = \mathscr{I}^A = \tilde{ \mathscr{I}}^A + \mu^{AB} \langle \O_B \rangle \; \;\; \mbox {or} \;\;\;\; \tilde{ \mathscr{I}}^A =  \mathscr{I}^A - \mu^{AB} \langle \O_B \rangle \label{shiftj}
\ee
The relation between the generating functionals is then 

\be
W_\mu [\tilde{\, \mathscr{I}}^A]= W[\,\mathscr{I}^A]-\frac{1}{2} \int \mu^{AB} \langle \O_A \rangle \langle \O_B \rangle 
\ee
In the above derivation,  it was assumed that the operator $\O_A$ is the same in the original and the deformed theory, and we used large $N$ factorization to effectively replace the $\O_A$ by their expectation values at the various steps. Note that while one adds $ S_{d.tr.}$ to the CFT action, one effectively subtracts it  from the generating functional, at least for the special case of a double-trace deformation.

Passing to holography, the generating functional $W[\,\mathscr{I}^A]$ is mapped to the renormalised on-shell action $S[\phi_A]$, which depends on the boundary values of the fields $\phi^A \leftrightarrow \mathscr{I}^A$, viewed as generalized coordinates. The operators $\O_A$ are to be identified with the generalized conjugate momenta, $\langle \O_A \rangle \leftrightarrow \pi_A $. The variation of the  on-shell action  as the sources are varied is

\be
\d S[\phi^A] = \int d^d x \, \pi_A \d \phi^A
\ee
The relation \eqref{shiftj} between the   sources in the deformed and undeformed theory can be simply reproduced by considering the variational principle for the total on-shell action in presence of the deformation.  Following the above discussion, we are instructed to subtract the functional of the expectation values $\mathcal{L}_{d.tr} (\O_A)$ from the generating functional, which amounts to performing a canonical transformation on the system. The variation of the total on-shell action is

\be
\d S_{tot} = \int d^d x    \, \left( \pi_A \delta \phi^A - \d \mathcal{L}_{d.tr} (\pi^A)\right)= \int d^d x \, \tilde \pi_A \d \tilde \phi^A  
\ee
For $\mathcal{L}_{d.tr.}$ given by \eqref{dtrl}, which only depends on the $\O_A$ but not the sources,  the effect of this canonical transformation is to shift the sources by an amount proportional to the expectation values

\be
\tilde \pi_A = \pi_A \;, \;\;\;\;\;\;\; \tilde \phi^A = \phi^A - \mu^{AB} \pi_B
\ee
while leaving the expectation values unchanged. This is of course equivalent, almost by definition, to the previous manipulations  at the level of the generating functional. 

The variational approach is useful for obtaining the deformed holographic data in more complicated situations, e.g. when the deformation depends on both the operators and the sources for them. A common situation occurs when one is interested in the expectation value of the stress tensor in the deformed theory. In this case, one should include the coupling of the deforming operator to a general background metric, \textit{i.e.}, a source for $T_{\a\b}$. As shown in \cite{Papadimitriou:2007sj}, the variational approach yields the correct shift in the expectation value of the stress tensor due to the deformation. This is a much easier computation than the direct manipulation of the generating functional. 

The situation we have at hand is even more complicated, because both the stress tensor and its source (\textit{i.e.}, the boundary metric or vielbein needed to covariantize the deformation) appear simultaneously in the deforming operator.  Thus,  we expect a change  both in the source and in the expectation value of the stress tensor as we perform the deformation. Having convinced ourselves that the variational approach should give equivalent results to the generating functional, we use it to greatly simplify the computation.

\subsection{Review of the $J\bar T$ deformation and its spectrum \label{jtrev}}

The $J\bar T$ deformation corresponds to a one-parameter family of two-dimensional QFTs, starting from a CFT,  which are related by the addition of the irrelevant $J\bar T$ operator to the euclidean action
\be
\frac{\p}{\p\mu} S(\mu)=  \int d^2 z  \, (J\bar T)_\mu \label{defjtbar}
\ee
where $J$ is a chiral $U(1)$ current and  $\bar T$ is the right-moving component of the stress tensor in the deformed theory; in principle, both of them depend on $\mu$. Using factorization of the $J\bar T$ operator inside energy-momentum-charge eigenstates, it is possible to derive an equation for how the energy levels  of the system placed on a cylinder of circumference $R$ evolve with $\mu$ \cite{Guica:2017lia} 

\be
\frac{\p}{\p\mu}  E(\mu,R) = 2 \int_0^R d\varphi \, \langle J \bar T \rangle = - Q \left(\frac{\p E}{\p R}   + \frac{P}{R} \right) \label{vare}
\ee
where the factor of $2$ accounts for the change of measure  from $z,\bar z$ to $\varphi, \tau$ coordinates, $d^2 z = 2 d\tau d\varphi$.
The momentum $P$ is quantized in units of $1/R$ and thus cannot vary with $\mu$. In \cite{Guica:2017lia}, it was assumed that also the charge $Q$ was quantized, and thus  $\mu$-independent. This led to a deformed energy spectrum of the form

\be
E_R = \frac{ 2\pi( h_R -\frac{c}{24})}{R - \mu Q_0} \;, \;\;\;\;\;\;\;\;\; E_L =  E_R + P = E_R + \frac{ 2\pi (h_L -  h_R)}{R} \label{oldspec}
\ee
where $E_{L,R} = \frac{1}{2} (E\pm P)$ are the left/right-moving energies in the deformed theory and  $h_{L,R}, Q_0$ are the left/right conformal dimensions and, respectively, the $U(1)$ charge  in the original CFT. 

However, if the chiral current $J$ is anomalous - which is a rather common situation - then the charge $Q$ can vary with $\mu$. A heuristic way to understand this is by viewing the $J\bar T$ deformation as a coupling between the chiral current $J$ and a  gauge field $\mathrm{a}_{\bar z} \propto  \bar T \, \d\mu$, as follows from \eqref{defjtbar}. If one computes the change in the divergence of $J$ due to the addition of the infinitesimal $J \bar T$ term to the  Lagrangian using conformal perturbation theory, one finds\footnote{The current $J$ above differs by a factor of $2\pi$ from the usual current $j=2\pi J$. The chiral anomaly coefficient $k$ is the one appearing in the $jj$ OPE, i.e.  $j(z) j(0) \sim k/2z^2$, and we assume it is constant along the flow.  }

\be
\d_\mu \, \bar \p J = \bar \p \int d^2 w J(z) J(w)\, \mathrm{a}_{\bar z} (w) = - \frac{k}{4\pi} \p \mathrm{a}_{\bar z}
\ee
Thus, it appears that the current $J$ is no longer conserved. However, it is possible to define a new instantaneous current $\hat J$
\be
\hat J_z = J \;, \;\;\;\;\;\; \hat J_{\bar z} = \frac{k}{4\pi} \mathrm{a}_{\bar z}
\ee
which is conserved, but no longer chiral. Written covariantly, this current reads
%
%

\be
\hat J^\a = J^\a +\frac{k}{8\pi}  (\g^{\a\b}+\e^{\a\b}) \, \mathrm{a}_\b \label{jhat}
\ee
 The charge $\hat Q$ associated with $\hat J$  is  equals the (quantized) charge $Q_0$ one had before the deformation. 

Of course, in our case we are not truly coupling to an external gauge field, and the current that enters the deformation is both chiral and conserved. A study of the related but much simpler $J\bar J$ deformation (see appendix \ref{cc}) suggests that the current $J$ that appears in the deforming operator is  obtained by restricting to the chiral component of the non-chiral conserved current $\hat J$. Since the charge associated with $\hat J$ is constant along the flow, it follows that the charge $Q$ associated with just its chiral component is not. More precisely

\be
\d_\mu \hat Q = \d_\mu \int_0^R \!\! d\varphi\,  (\hat J_z - \hat J_{\bar z}) = \d_\mu Q - \frac{k R}{4 \pi} \,\mathrm{a}_{\bar z} =0 
\ee
Plugging in the instantaneous value for $\mathrm{a}_{\bar z}$, we find

\be
\frac{\p}{\p \mu}Q = 
 \frac{k}{4\pi} R \,\langle T_{\bar z \bar z} \rangle = - \frac{k}{4\pi} R \, \p_R E_R \label{flowq}
\ee
which turns out to be the correct flow equation obeyed by the charge of the $U(1)$ current. Equations \eqref{vare} and \eqref{flowq} can then be combined to find out the full spectrum $E_{L,R}(\mu, R)$ and $Q (\mu, R)$ quoted below. 

To make the above arguments rigorous, we would need a better understanding of how the current behaves in conformal perturbation theory, and in particular of our assumption that $J$ equals the chiral part of $\hat J$ at any point along the flow. There is, however, a more rigorous indirect argument for obtaining the modified spectrum, given in  \cite{Chakraborty:2018vja}.  The argument  is based upon splitting the left-moving stress tensor into a  contribution from the chiral current, which takes the Sugawara form, and an independent ``coset'' contribution, whose OPE with the current vanishes along the flow. \cite{Chakraborty:2018vja} then argue that the coset part is unaffected by the deformation, which implies a spectral-flow type equation for the left-moving energy

\be
E_L R - \frac{2\pi Q^2}{k} = const = 2\pi \left(h_L - \frac{c}{24} - \frac{Q_0^2}{k} \right)\label{model}
\ee 
Using constancy of $P = E_L - E_R$, one finds that $E_R R =2\pi Q^2/k + const$. Combining this with \eqref{vare}, it is possible to show that that $Q$ precisely satisfies  the equation \eqref{flowq} above. Trading the $R$ derivative for a $\mu$ derivative, one finds that the charge varies along the flow as

\be
Q= Q_0 + \frac{\mu k}{4\pi} E_R \label{newq}
\ee
The solution for the left/right-moving energies in terms of the original CFT data is

\be
E_R = \frac{4\pi}{\mu^2 k} \left( R - \mu Q_0 - \sqrt{\left(R-\mu Q_0 \right)^2 - \mu^2 k \left( h_R - \frac{c}{24} \right)}\right) \;, \;\;\;\;\;\; E_L = E_R + \frac{2\pi (h_L - h_R)}{R} \label{moder}
\ee
It is easy to check that in the $k\r 0$ limit, these equations reduce to \eqref{oldspec}. 
It is also amusing to note that the equation for the right-moving energy can be suggestively written as  a spectral flow equation 

\be
E_R \left(R-\mu Q_0\right) - \frac{2\pi}{k} \left( \frac{\mu k E_R}{4\pi} \right)^2 = const = 2\pi \left(h_R - \frac{c}{24} \right) \label{specfler}
\ee 
if we interpret $ -\mu k E_R/4 \pi$ as a right-moving contribution to the $U(1)$ charge, such that the total charge  $\hat Q = Q + Q_R = Q_0 = const$ along the flow. The first term is the effective shrinking of the radius seen by the right-movers, which was already visible in absence of the anomaly in \eqref{oldspec}. 

One of the main results of this paper is to reproduce the $k=0$ and $k\neq 0$ spectra \eqref{oldspec} and, respectively,  \eqref{moder} from holography. Since, as noted above, we do not have a full understanding of how the chiral current $J$ behaves along the flow when the chiral anomaly is present, our most rigorous results are for the $k=0$ case. However,  we find that a simple modification
 of  the dictionary we derive (which is anomaly blind)  allows us to correctly reproduce the spectrum  also in presence of the  anomaly. This modification is consistent with our heuristic picture for  how the charge should  behave along the flow.

\subsection{Sources and expectation values in the $J\bar T$-deformed theory}

 The minimal set of phase space variables that we need to consider are the stress tensor $T^a{}_\a$ and current $J^\a$, which are canonically conjugate to the boundary vielbein $e^\a{}_a$ and gauge field, $\mathrm{a}_\a$.  Here, latin indices denote the tangent space and greek ones are spacetime indices. The reason  we prefer the vielbein formulation is that the deformed theory is not Lorentz invariant;  consequently, the conserved stress tensor is not  symmetric and it  naturally couples to the vielbein, and not the metric. The variation of the original CFT action reads 

\be
\d S_{CFT} = \int d^2 x\, e \, \left( T^a{}_{\a}\, \d e^{\a}_a + J^\a \d \mathrm{a}_\a \right) \label{varcftact}
\ee
where $e = \det e^a_\a$ and from now on we will omit the brackets from the expectation values of the various operators.

Next, we add the double-trace $J\bar T$ deformation, appropriately covariantized. Since the coupling parameter is a dimensionful null vector, in  an arbitrary background it makes the most sense to keep the coupling with tangent space indices fixed, so $\mu^a= \mu \, \d^a_-$, where $\mu$ is a constant with dimensions of length and $x^-$ is a null direction along the boundary.  The covariantized multitrace operator is then

\be
S_{J\bar T} = \int d^2 x \, e \, \mu_a T^a_\a J^\a
\ee
The problem that we would like to solve is to find new canonical variables such that the variation of the total action, including the multitrace contribution, can be written in the form   $\tilde \pi_A \d \tilde \phi^A$ for some new canonical variables $\tilde \pi_A$, $ \tilde \phi^A$. That this should be possible is guaranteed by the fact that we are performing a canonical transformation. The variation of the action including the multitrace is
\bea
\d S - \d S_{J\bar T}& = & \int d^2 x \left[ e\, T^a_\a \d e^\a_a + e \,J^\a \d \mathrm{a}_\a - \d (e\, \mu_a T^a_\a J^\a)\right]\label{vartotact}  \\
&=& \int d^2 x \, e \left[ \, T^a_\a (\d e^\a_a- \mu_a \d J^\a) + J^\a (\d \mathrm{a}_\a -\mu_a \d T^a _\a) + e^a_\a \d e^\a_a \, \mu_b T^b_\b J^\b\right] \nonumber\\ 
&=& \int d^2 x \, e \left[ \,( T^a_\a + e^a_\a \, \mu_b T^b_\b J^\b) (\d e^\a_a- \mu_a \d J^\a) + J^\a (\d \mathrm{a}_\a -\mu_a \d T^a _\a) +\mu_a e^a_\a \d J^\a \, \mu_b T^b_\b J^\b   \right] \nonumber
\eea
To proceed, we use the fact that $J$ is purely chiral $J = J_+ (x^+)$, and so $\mu_a J^a=0$. The above expression can then be  manipulated into
\be
\d S - \d S_{J\bar T}= \int d^2 x \, e \left[ \,( T^a_\a + ( e^a_\a + \mu_\a J^a) \, \mu_b T^b_\b J^\b) (\d e^\a_a- \mu_a \d J^\a) + J^\a (\d \mathrm{a}_\a -\mu_a \d T^a _\a)   \right]  \label{newsv}
\ee
We can easily read off the modified sources and expectation values from the above expression\footnote{Note that the first equation and the condition $\mu_a J^a =0$ imply that $\tilde e = e$ and $\tilde{e}^a_\a = e^a_\a + \mu_\a J^a$.}

\setlength{\jot}{8pt}
\medskip
\begin{empheq}[box=\widefbox]{align}
\tilde e^{\a}_a  = e^{\a}_a - \mu_a J^\a \;, \;\;\;\;\; \tilde{\mathrm{a}}_\a  = \mathrm{a}_\a^{} -\mu_a T^{a}_\a \;\;\;\;\; \nonumber 
\\ 
\tilde T^a_\a  = T^{a}_\a + (\mu_b T^{b}_\b J^\b)\, \left(e^{a }_\a  +  \mu_\a J^a \right) \;, \;\;\;\;\; \tilde J^\a  = J^{\a} \label{newvevs}
\end{empheq} 

\setlength{\jot}{2pt}
\vskip2mm

%
\noindent This is one of our main results. The tilded quantities are evaluated in the deformed theory with parameter $\mu_a$, whereas the untilded quantities belong to the original CFT at $\mu=0$. We have defined $\mu_\a = \mu_a e^a_\a$ and $J^a = e^a_\a J^\a$.  Note that the expression for the sources coincides with the naive expression  \eqref{shiftj}, for  $\mu_{AB}$  off-diagonal.

As mentioned in the previous subsection, we do not fully understand yet how to relate the chiral $U(1)$ current in the deformed theory to  the original current when a chiral anomaly is present. We can thus only be certain of the correctness of our derivation for $k=0$. However, as we will show in section \ref{matft}, a holographic dictionary that is identical to \eqref{newvevs} except for the last equation  reproduces the correct spectrum \eqref{moder} also for $k$ nonzero, indicating that the corrections  due to the chiral anomaly are contained in the changes to the current.

One possible objection to our  manipulations  above is that, according to its definition, the  $J \bar T$ deformation is supposed to involve the instantaneous stress tensor in the theory deformed by $\mu$, and not the stress tensor of the original CFT, as written in \eqref{vartotact}. However, it is easy to check that  $\tilde e\, \mu_a \tilde T^a_\a \tilde J^\a = e\, \mu_a T^a_\a J^\a$, so this distinction does not matter for our derivation. Alternatively, one could implement the instantaneous deformation by considering the 
 infinitesimal version of the equations  \eqref{newvevs}, obtained from the last line of \eqref{vartotact} by discarding the $\O(\mu^2)$ term, and then integrating with respect to $\mu$. While one naively obtains different expressions\footnote{Naively integrating one obtains $\tilde T^b_\b  = T^{b}_\b + (\mu_a T^{a}_\a J^\a)\, \left(e^{b }_\b  + \frac{1}{2} \mu_\b J^b \right)$ and $ \tilde{\mathrm{a}}_\b = \mathrm{a}_\b^{} -\mu_b T^{b}_\b-\frac{1}{2}  (\mu_a T^{a}_\a J^\a)\,\mu_b e^{b }_\b $.} for $\tilde{\mathrm{a}}_\a$ and $\tilde T^a_\a$, it is possible to show that the extra term in $\tilde{\mathrm{a}}_\a$ can be moved to $\tilde T^a_\a$ without affecting the variational principle. This freedom is due to the fact  that \eqref{vartotact}  does not uniquely determine the flow equations for the field theory data, but some additional criteria are needed, such as requiring that the deformed expectation values satisfy the appropriate Ward identities.

\subsection{Ward identities}

We end with a note on the Ward identities satisfied by the deformed stress tensor and the current, which will be useful in the later sections. The Ward identities in the original CFT can be obtained from the variation of the CFT action \eqref{varcftact},
specialized to gauge transformations and diffeomorphisms, for which $\d S_{CFT} =0$. 
Invariance under gauge transformations $\mathrm{a}_\a \r \mathrm{a}_\a + \p_\a \l $ imposes that $\nabla_\a J^\a =0$, though for $J$ chiral there will generally be an anomaly. 
 Invariance under diffeomorphisms\footnote{We are assuming that $c_L = c_R = c$, so there is no gravitational anomaly on the boundary. }, under which
\be
\d_\xi e^\a_a = - \xi^\l \p_\l e^\a_a + \p_\l \xi^\a e^\l_a \;, \;\;\;\;\; \d_\xi \mathrm{a}_\a = -\xi^\l \mathrm{f}_{\l\a} - \p_\a (\xi^\l \mathrm{a}_\l)
\ee
implies the conservation equation 

\be
\nabla_\b (T^a_\a e^\b_a) + T^a_\b \nabla_\a e^\b_a - T_{ab} \, \om_\a{}^{ab}+ \mathrm{f}_{\a\b} J^\b - \mathrm{a}_\a \nabla_\b J^\b =0
\ee
The second term vanishes by the tetrad postulate, the third when $T_{ab}$ is symmetric, the fourth if the gauge connection is flat and so does the last one, using current conservation. Thus, we find that the stress tensor is conserved for any boundary metric, as long as it is symmetric and $\mathrm{f}_{\a\b}=0$. 

 After adding the deformation, the variation of the action  is given by \eqref{newsv}. In terms of the new ``tilded'' variables defined in 
 \eqref{newvevs}, the Ward identity is identical with the one above

\be
\tilde \nabla_\l (\tilde T^a_\a \tilde e^\l_a ) + \tilde  T^a_\l \tilde \nabla_\a \tilde e^\l_a  - \tilde T_{ab} \, \tilde{\om}_\a{}^{ab}+ \tilde{\mathrm{f}}_{\a\b}  J^\b - \tilde{\mathrm{a}}_\a \tilde \nabla_\l J^\l =0 \label{newwardid}
\ee
If we assume that the original $\mathrm{f}_{\a\b} = \nabla_\a \mathrm{a}^\a= 0$, then $\tilde e = e$ implies that $\tilde \nabla_\a J^\a =\nabla_\a J^\a = 0$. The second  term can be dropped as before, while the fourth term $\tilde{\mathrm{f}}_{\a\b}=\mathrm{f}_{\a\b} - \mu_a (\p_\a T^a_\b - \p_\b T^a_\a)$ will vanish provided that $\mu_a (\p_\a T^a_\b - \p_\b T^a_\a)=0$. This will indeed be the case for the backgrounds we will consider.  Thus, the new stress tensor (which in general will not be symmetric) will be conserved with respect to the new background $\tilde e_{\a}^a$, provided the spin connection vanishes.

\section{The holographic dictionary \label{holodict}}

The analysis of the previous section reveals a very simple way to construct the holographic dictionary for the $J\bar T$-deformed CFT, starting from the usual AdS$_3$/CFT$_2$ holographic dictionary. Namely, the holographic dictionary in presence of sources $\tilde e^a_\a$, $\tilde{\mathrm{a}}_\a$ for the stress tensor and the current in the deformed theory can be constructed in two steps:

\begin{itemize}
\item[i)] First, one works out the usual AdS/CFT dictionary in presence of the boundary sources 
\be
e^\a_a = \tilde e^\a_a + \mu_a J^\a \;, \;\;\;\;\; \mathrm{a}_\a = \tilde{\mathrm{a}}_\a + \mu_a T^a_\a  \label{defolds}
\ee
where $\tilde e^a_\a$, $\tilde{\mathrm{a}}_\a$ are held fixed and  $  T^a_\a $ and $ J^\a $ are determined by holographically computing the expectation values of the CFT stress tensor and current in the above background and feeding them back into the sources
\item[ii)] To find the expectation values  $  \tilde{T}^a_\a $, $ \tilde J^\a $ in the deformed theory, one simply plugs in the  values of  $T^a_\a $, $ J^\a $  found at step i) into \eqref{newvevs}.
\end{itemize}

\noindent The dual geometry will simply be given by the asymptotically locally AdS$_3$ solution that obeys the boundary conditions \eqref{defolds}. Note that the only role of holography in this procedure is to provide a simple means to  compute the holographic expectation values in the undeformed CFT with non-trivial boundary sources; in particular, step ii) of the procedure is purely field-theoretical. 

To carry out step i),  all we need is  the  usual AdS$_3$/CFT$_2$ dictionary for three-dimensional Einstein gravity  coupled to   $U(1)$ Chern-Simons, in presence of arbitrary boundary sources. Both of these dictionaries are extremely well-studied \cite{Balasubramanian:1999re,deHaro:2000vlm,Kraus:2006nb}.   
The chiral anomaly is easily visible in holography, as it appears classically at the level of the Chern-Simons term. However,  as explained above, so far we  are only confident about our understanding of the holographic dictionary for the $k=0$ case. Consequently, the analysis of the rest of the paper will be split into two cases that we treat separately, namely $k=0$ and $k \neq  0$.

For $k=0$, the holographic dictionary is fully understood (as $J$ does not shift); however,  the contribution of the Chern-Simons term is a bit subtle in this limit, and needs to be treated with care. We do this in section \ref{fermi}
, finding results that are in perfect agreement with \eqref{oldspec} $k=0$.  For $k\neq 0$, the Chern-Simons contribution is entirely standard, but we do not have a clear understanding of how to relate the chiral current in the deformed theory to its undeformed counterpart; despite this, we will be able to match the spectrum also in this case after making one natural assumption  about the expectation value of the current.

We  start this section by reviewing the holographic dictionary for AdS$_3$ with Dirichlet boundary conditions, in presence of arbitrary boundary sources. In \ref{asyjtbar}, we write down the most general asymptotic solution that corresponds to the deformed theory with all sources set to zero. In \ref{fermi}, we compute the holographic expectation value of the stress tensor for $k=0$, which is somewhat subtle, by invoking the expected equivalence between $U(1)$ Chern-Simons and a pair of chiral fermions. In \ref{holovevs} we present the holographic one-point function of the stress tensor for the case $k \neq 0$.

\subsection{Review of the AdS$_3$/CFT$_2$ holographic dictionary \label{revadscft}}

We review herein the standard AdS$_3$/CFT$_2$ dictionary, namely with Dirichlet boundary conditions on the fields. In the bulk, our system consists of Einstein gravity with a negative cosmological constant coupled to a $U(1)$ Chern-Simons gauge field,

\be
S_{bulk}= \int d^3 x \sqrt{g} \left[ \frac{1}{16 \pi G} \left(R + \frac{2}{\ell^2} \right) + \frac{k}{8\pi} \, \e^{\mu\nu\rho} A_\mu \p_\nu A_\rho \right] \label{sbulk}
\ee
The most general solution for the metric in radial gauge is given by the usual Fefferman-Graham expansion 

\be
ds^2 = \ell^2 \,\frac{dz^2}{z^2} + \left(\frac{g^{(0)}_{\a\b}}{z^2} + g^{(2)}_{\a\b} + z^2 g^{(4)}_{\a\b}\right) dx^\a dx^\b
\ee
which terminates in three dimensions \cite{Skenderis:1999nb}. The boundary metric, $g^{(0)}$, is arbitrary; the asymptotic equations of motion fix the trace and divergence of $g^{(2)}$ in terms of $g^{(0)}$, which will yield the holographic Ward identities; the component $g^{(4)}$ is entirely determined by the previous two. As for the gauge field, the Chern-Simons equations of motion require that the connection be flat. In radial gauge ($A_z=0$), the most general solution for $A_\mu$ is then

\be
A = A_\a (x^\a) dx^\a \;, \;\;\;\;\; dA=0
\ee
where the $A_\a$, with $\a = \pm$ (using null coordinates on the boundary), are $z$-independent.

The first step in finding the holographic dictionary is to ensure that the variational principle - in this case, with Dirichlet boundary conditions at the AdS$_3$ boundary - is well defined. For this,  the gravitational bulk action  \eqref{sbulk} needs to be supplemented by a Gibbons-Hawking boundary term

\be
S_{bnd, D}^{grav}=  \frac{1}{8 \pi G} \int d^2 x \sqrt{\g}   \, K 
\ee
An additional  boundary counterterm $S_{ct} =- \frac{1}{8 \pi G \ell}\int d^2 x \sqrt{\g}$ is needed to render the expectation value of the holographic stress tensor finite. The variation of the total gravitational on-shell action is then\footnote{Note the sign difference with respect to \eqref{varcftact} in the definition of the stress tensor. This  is due to the difference of definitions of the stress tensor in Euclidean versus Lorentzian signature, which follows from $- S_E = i S_L$ and $\tau_E = i t_L$. The expectation values of the two stress tensors are nevertheless the same, after analytic continuation of the time coordinate. The same comments apply to the variation of the Chern-Simons action below.}

\be
\d S_{tot}^{grav} =\frac{1}{16 \pi G} \int_{z=\e} d^2 x \sqrt{\g} \,\left( K_{\a\b} - \g_{\a\b} \, K + \frac{1}{\ell}\,\g_{\a\b}\right) \d \g^{\a\b}=  - \int d^2 x \sqrt{g_{(0)}} \,\frac{1}{2} \, T_{\a\b}^{grav} \d g_{(0)}^{\a\b}
\ee
Plugging in the explicit expression for the extrinsic curvature in terms of the asymptotic expansion for the metric, we find
\be
T^{grav}_{\a\b}  = \frac{1}{8 \pi G \ell}\, \left(   g^{(2)}_{\a\b} - g^{(0)}_{\a\b} \frac{\ell^2}{2} \, R^{(0)} \right) \label{holost}
\ee
The equations obeyed by $g^{(2)}$ ensure that $T^{(grav)}$ obeys the holographic Ward identities, \textit{i.e.}, it is conserved with respect to  the boundary metric for any $g^{(0)}$, and its trace is in agreement with the holographic conformal anomaly.

The usual treatment of the  Chern-Simons term is as follows (see e.g. \cite{Kraus:2006wn}).  Plugging in the asymptotic expansion of the gauge field into the on-shell variation of the Chern-Simons action, one notes that the two components $A_\pm$ of the gauge field on the boundary are canonically conjugate to each other. Therefore, in order for the  variational principle to be well-defined,  only one of them can be fixed. By adding the counterterm 
\be
S_{bnd}^{\text{\tiny{CS}}} = \mp \frac{ k}{16 \pi}  \int d^2 x \sqrt{\g} \, \g^{\mu\nu} A_\mu A_\nu \label{csct}
\ee
one  has $\d S_{CS} \propto \d A_\mp$; the upper sign corresponds to fixing $A_-$ on the boundary, and the lower one to fixing $A_+$. From now on, we will choose the upper sign, which yields to a chiral (as opposed to an antichiral) boundary current. The variation of the total on-shell Chern-Simons action takes the form\footnote{Note again this current differs by a factor of $2\pi$ from the usual definition.} 
\be
\d S^{\text{\tiny{CS}}}_{tot} = -\int d^2 x \sqrt{\g} \, \left(\frac{1}{2} \, T_{\a\b}^{\text{\tiny{CS}}} \d \g^{\a\b} +   J^\a \d A_\a \right)
\ee
where the contribution to the stress tensor is due to explicit dependence of the  counterterm \eqref{csct}  on the boundary metric. The expectation value of the current is given by  

\be
J^\a =  \frac{k}{8\pi} (A^\a - \e^{\a\b} A_\b) \label{curr}
\ee
while 
\be
T_{\a\b}^{\mathrm{c}\mathrm{s}} =  \frac{k}{8\pi} (A_\a A_\b - \frac{1}{2} \g_{\a\b} A^2) \label{tcsnaive}
\ee

\subsection{Asymptotic expansion dual to the $J\bar T$-deformed theory \label{asyjtbar}}

Starting from this section, we will fix zero boundary sources in the deformed theory, namely we take $\tilde e^\b_{b}  = \d^\b_b$ and  $\tilde{\mathrm{a}}_\b =0$. 
This will allow us to compute general one-point functions in the deformed theory, but no higher-point correlators. 
Using the dictionary \eqref{defolds}, the asymptotic expansion of the dual bulk fields in  AdS$_3$  will be given by the usual Fefferman-Graham expansion for the particular case when  the boundary  sources take the form 

\be
e^{\a } _a = \d^\a_a + \mu_a  J^\a  \;, \;\;\;\;\;\;\mathrm{a}_\a =\mu_a   T^a{}_{\a} \label{afbndc}
\ee
At the level of the metric, the new boundary condition corresponds to

\be
g^{(0)}_{\a\b} = \eta_{\a\b} - \mu_\a  J_\b - \mu_\b J_\a \label{modmet}
\ee
or, in components\footnote{There is a single non-vanishing Christoffel symbol, $\Gamma^-_{++} = P''(x^+)$ , the boundary Ricci scalar is  zero and the boundary vielbeine read \be
e^a{}_\a = \left( \begin{array}{cc} 1 & 0 \\ P' & 1\end{array}\right) \;, \;\;\;\;\;\;e_{a\a} = \frac{1}{2} \left( \begin{array}{cc} P' & 1 \\ 1 & 0\end{array}\right)  \nonumber
\ee
The associated spin connection is $\om_\a{}^a{}_b = - e^\l_b \nabla_\a e^a_\l =0$. }
\be
g^{(0)}_{++} = -  \mu J(x^+) \equiv  P'(x^+)\;, \;\;\;\;\;\; g^{(0)}_{+-} = \frac{1}{2} 
\ee
Above, we have made use of the assumption that the current $J^\a$ is purely chiral, so $J^+=0$ and $J^- = 2 J_+ = 2 J(x^+)$.  The quantity $P(x^+)$ has been introduced in order to make contact with the notation of \cite{Compere:2013bya}, who considered the same boundary metric. We will oftentimes use $P'$ instead of $- \mu J$ throughout the text.  Note that for general $J(x^+)$, this metric is induced by the coordinate transformation

\be
x^+ \r x^+ \;, \;\;\;\;\;\; x^- \r x^- - \mu \int J(x^+) \, dx^+ = x^- + P(x^+) \label{coordtr}
\ee
The most general solution to Einstein's equations satisfying these boundary conditions has

\be
g^{(2)}_{++} = \mathcal{L}(x^+) + \bar \L(x^- + P(x^+)) P'(x^+)^2 \;, \;\;\;\;\;\; g^{(2)}_{+-} = \bar \L(x^- + P(x^+)) P'(x^+) \label{asymet}
\ee

\be 
g^{(2)}_{--} = \bar \L(x^- + P(x^+))
\ee
where $\L,\bar \L$ are two arbitrary functions of their respective arguments. Again,
this can be simply obtained by applying   the above coordinate transformation to the most general asymptotically AdS$_3$ solution with Dirichlet boundary conditions.

The functions $\mathcal{L}$, $\bar{\mathcal{L}}$ and $P$ entirely determine the gravitational solution. They also entirely determine the solution for the gauge field, if we assume the latter can be decomposed into a part $\mathrm{a}_\a $ that represents the boundary source and a part $A_\a^{vev}$ that is proportional to expectation value   of the CFT current
\be
A_\a =  \mathrm{a}_\a  +A_\a^{vev} \label{adec}
\ee
The source $\mathrm{a}_\a $ is determined by the boundary condition \eqref{afbndc}, while $A^{vev}_\a$ is determined through the holographic relation  \eqref{curr}, which in this background reads

\be
 \mathcal{J}^- = 2 \mathcal{J}(x^+) = \frac{k}{2\pi} (A_+ - P'(x^+) A_-) 
\ee
According to the holographic dictionary \eqref{newvevs}, we should have $ \mathcal{J}(x^+)=J(x^+)$, but we use this slightly different notation for reasons that will become clear later. Current conservation and the Chern-Simons equation of motion, $F_{\mu\nu} =0$, require that
 
\be
\p_- A_+ = P'(x^+) \p_- A_- = \p_+ A_-
\ee  
which implies that
\be
A_- = \mathcal{A}(x^- +P(x^+)) \;, \;\;\;\;\;\; A_+ = \frac{4\pi}{k}  \mathcal{J}(x^+) + P'(x^+)  \mathcal{A} (x^- +P(x^+))  \label{formgf}
\ee
for some function $\mathcal{A}$. Using the decomposition \eqref{adec}, the terms proportional to $\mathcal{A}$ must correspond to the gauge field source, whose components read

\be
 \mathrm{a}_+ = P'  \mathcal{A}\;, \;\;\;\;\;\;  \mathrm{a}_-=  \mathcal{A} \label{extgf}
\ee
These will  be identified with the stress tensor via \eqref{afbndc} or, more precisely,

\be
 \mathrm{a}_\pm =  \mu \, T_{-\pm}  \label{atrel}
\ee 
which will in turn determine $\mathcal{A}$ in terms of $\bar\L$.
Note that by itself, the boundary source satisfies $\mathrm{f}_{\a\b} = \nabla_\a \mathrm{a}^\a=0$.
 This is consistent with \eqref{afbndc} provided that $\mu_a \p_{[\a} T^a_{\b]}=0 $, which we checked. 

To find the solution for $\mathcal{A}$, we need to compute $T_{a\a}  = e_a^{\b}  T_{\a\b} $. The stress tensor is the sum of   a gravitational and a Chern-Simons contribution

\be
T_{\a\b} = T^{grav}_{\a\b}+T^{\text{\tiny{CS}}}_{\a\b} \label{ttot}
\ee
which are given in \eqref{holost} and \eqref{tcsnaive}. The stress tensor with one tangent space and one spacetime index is computed as

\be
T_{a\a} = e^{\b}_{a} T_{\b\a}
= \frac{1}{8\pi G \ell} \left( T_{\b\a} \d^\b_a + \mu_a T_{\a\b} J^\b\right)
\ee
or, in components

\be
T_{++}
= \frac{\L}{8\pi G \ell} + \frac{2\pi}{k} \mathcal{J}^2\;, \;\;\;\;\; T_{+-} = 
0 \nonumber
\ee

\be
T_{-+} = \frac{\bar \L \, P'}{8\pi G \ell} + \frac{k}{8\pi} \mathcal{A}^2 P'\;, \;\;\;\;\;\; T_{--} = \frac{\bar \L}{8\pi G \ell} + \frac{k}{8\pi} \mathcal{A}^2 \label{expt0}
\ee
Note that the  coordinate dependence of the stress tensor components and their ratio is consistent with the  coordinate dependence and ratio of $\mathrm{a}_\pm$.

In order to finalize writing the bulk solution, we need to relate the functions $\mathcal{A}$ and $\mathcal{J}$ to the other functions appearing in the metric and the gauge field, as instructed by the holographic dictionary. We will separately consider the cases $k=0$ and $k \neq 0$. 

\subsubsection*{\bf{Chern-Simons level} $k=0$}

When the Chern-Simons level $k\r 0$, we can neglect the Chern-Simons contribution to $T_{-\pm}$ in \eqref{expt0}.  Consequently, the relation \eqref{atrel} between the function $\mathcal{A}$ determining the gauge field source and the expectation value of the stress tensor becomes

\be
\mathcal{A}(x^- + P(x^+))  = \frac{\mu \bar{\mathcal{L}}(x^- + P(x^+)) }{8 \pi G \ell} \label{abulkkz}
\ee
We also use the holographic dictionary \eqref{newvevs}, which we trust at $k=0$, to relate

\be
\mathcal{J}(x^+) = J(x^+)
\ee
in the solution \eqref{formgf} for the gauge field. With this, the bulk solution is completely specified. Note that it is parametrized by as many free functions as in the case of Dirichlet boundary conditions. In fact, the full bulk solution can be obtained by applying the field-dependent coordinate transformation \eqref{coordtr}, together with a field-dependent gauge transformation

\be
\l = \mu \int dx^- T_{--} (x^- + P(x^+)) \label{gaugetrsol}
\ee
to the most general AdS$_3$ solution with Dirichlet boundary conditions and zero gauge field source. 
Note that both transformations break the fields' periodicities.


\subsubsection*{\bf{Non-zero Chern-Simons level}}

For non-zero Chern-Simons level,  we find that the gauge source $\mathrm{a}_-= \mathcal{A}$ satisfies a quadratic equation

\be
 \mathcal{A} = \frac{\mu\bar{\mathcal{L}}}{8 \pi G \ell} +  \frac{\mu k}{8\pi} \mathcal{A}^2 \label{eqA}
\ee
which follows from \eqref{atrel} and \eqref{expt0}. The solution that is smooth as $\mu \r 0$ is

\be
\mathcal{A} = \frac{4\pi}{k \mu} \left(1- \sqrt{1- \frac{\mu^2 k \bar{\mathcal{L}}}{16 \pi^2 G \ell}} \right) \label{solA}
\ee
Note that the gauge field becomes imaginary when $\bar{\mathcal{L}}$ is too large, which is reminiscent of the upper bound on the right-moving conformal dimension found in the field theory analysis \eqref{moder}. To fully specify the bulk solution, we also need to spell out the relation between $\mathcal{J}$ and $J$. As we will see in section \ref{matft}, when the expectation values are constant we are able to match the finite $k$ spectrum by assuming that $\mathcal{J}$ takes a particular value we can justify. However, a better, first-principles  understanding   of the holographic dictionary for $k \neq 0$ is necessary.

\subsection{The holographic expectation values for $k=0$ \label{fermi}}

We start with the simpler $k=0$ case, in which the current $J$ is unchanged along the flow. As explained, in this case the Chern-Simons contribution to the antiholomorphic stress tensor in \eqref{expt0} can be dropped, leading to the simplified bulk solution \eqref{abulkkz}. However, the Chern-Simons contribution  to the holomorphic stress tensor appears  to diverge in this limit, so it must be treated with care. 

To understand what to expect, it is useful to take the $k \r 0$ limit of the exact spectrum formula \eqref{model}, in which the Chern-Simons contribution should roughly be identified with the $2\pi Q^2/k$ shift.  As $k\r 0$, $E_L$ receives a  contribution proportional to $2\pi Q_0^2/k$, which diverges, but is the same as the corresponding contribution in the undeformed CFT. The \emph{change} in $E_L$ with respect to its CFT value is thus finite as $k\r0$, and from \eqref{model}, \eqref{newq} we find that it is proportional to $\mu  E_R$. 

One way to obtain the current sector's contribution to the stress tensor as $k\r 0$ without using the dependence of the current on $k$ 
 is to model  the Chern-Simons term by a pair of chiral fermions and treat them ``classically'', i.e. ignoring the effects of the anomaly. The action for fermions coupled to an external  gauge field $\mathrm{a}_\a $ is

\be
S = \int d^2 x \, e \, i \bar{\Psi} \g^a e_a^\a (\mathcal{D}_\a - i \mathrm{a}_\a) \Psi \label{fermact}
\ee
where $\mathcal{D}_\a$ includes the spin connection and the fermions are chiral,  $\g^3 \Psi = \Psi$.   We are interested in computing the stress tensor defined in \eqref{varcftact}, which couples to the vielbein $e^\a_a$. The result should then be evaluated on the background \eqref{afbndc}, in presence of the pure gauge source \eqref{extgf}, and should yield the Chern-Simons contribution to the stress-tensor one-point function in \eqref{ttot} as $k \r 0$.

This stress tensor is  obtained by varying the  action \eqref{fermact} with respect to the vielbein; however, in order to better understand how the presence of the gauge field affects the physical left-moving solution for the fermions, we will plug in the background \eqref{afbndc} directly into the action and compute the stress tensor via the canonical method. Due to the explicit time dependence of the gauge field, the latter is not conserved; however, it is related to the gauge-invariant stress tensor via 

\be
T^{gauge\,inv.}_{a\a} = T^{can}_{a\a} + J_a a_\a 
\ee
which is conserved because $\mathrm{f}_{\a\b} =0$. This gauge-invariant stress tensor coincides with the one obtained by varying with respect to the vielbein. Plugging in the explicit two-dimensional gamma matrices and $\Psi = (\psi\;  0)$, the action simplifies to
\be
S = -i \int d^2 x \,  \psi^\star ( \p_- -i \mathrm{a}_{-}) \psi
\ee
The equation of motion is  $\p_- \psi = i \mathrm{a}_{-} \psi$, with solution

\be
\psi (x^+,x^-) = e^{i \l (x^+,x^-)} \psi^{(0)} (x^+) \;, \;\;\;\;\; \p_- \l = \mathrm{a}_{-} \label{psieom}
\ee
where $\psi^{(0)}$ is the solution in absence of the gauge field, which is purely left-moving. 
The only non-zero component of the gauge-invariant stress tensor $T_{a\a}$, evaluated on the solution, is 

\be
T_{++} = - \frac{i}{2} \psi^\star ( \p_+ - i \mathrm{a}_+) \psi  = - \frac{i}{2} \psi^{(0)\star}  \p_+ \psi^{(0)} - \frac{1}{2} \psi^\star \psi\, (\mathrm{a}_+ - \p_+ \l)
\ee
Note that even though the external source is pure gauge, $\mathrm{a}_+$ need not exactly equal $\p_+ \lambda$, since the gauge parameter is constrained to respect the fermion's periodicity condition, i.e. Neveu-Schwarz or Ramond\footnote{Note that the parameter $\l$ in \eqref{psieom} is \emph{not} the same as the one in \eqref{gaugetrsol}, but they differ at the level of winding modes. }. Rewriting the above equation in terms of the undeformed stress tensor $T^{(0)}_{++}$ and the current, the change in the stress tensor due to adding the external source is

\be
T_{++} = T^{(0)}_{++} + J(x^+) ( \mathrm{a}_+ - \p_+ \l )
\ee
where the normalizations have been fixed by requiring that they map to the standard normalization for a chiral boson. More generally, one would also expect contributions to the stress tensor that are quadratic  in the  gauge field \cite{Treiman:1986ep}
; however, these terms vanish in our case because we have set to zero the coefficient of the chiral anomaly. 

 We would now like to evaluate the expectation value of the above stress tensor on a cylinder of circumference $R$, when 
the gauge field to which we are coupling  has components

\be
\mathrm{a}_- = \frac{\mu \bar{\mathcal{L}} (x^- + P(x^+))}{8\pi G\ell} \;, \;\;\;\;\; \mathrm{a}_+ = \frac{ \mu  P' \bar{\mathcal{L}} (x^- + P(x^+)) }{8\pi G\ell}
\ee
This form of the gauge field follows from \eqref{atrel}, after using the above-derived fact that the current contribution to $T_{-\pm}$ is zero. 
Expanding the function $ \bar{\mathcal{L}} (x^- + P(x^+)) $ in Fourier modes $\bar{\mathcal{L}}_n$, the solution for $\l$ defined in \eqref{psieom} is

\be
\l (x^+,x^-) =\frac{\mu}{8\pi G\ell} \left[ \bar{\mathcal{L}}_0 (x^- - x^+) - \frac{i}{2\pi}   R' \sum_{n \neq 0} \frac{\bar{\mathcal{L}}_n }{ n}\, e^{\frac{2\pi i n }{R'}(x^- +P(x^+))} \right]
\ee 
where, importantly, the first term multiplying the $\bar{\mathcal{L}}$ zero mode is fixed by requiring that $\l$ have no winding mode, which is the same as requiring the correct periodicity of the deformed fermion, and $ R'$ is the periodicity of the coordinate $x^- + P(x^+)$. 
 It is then easy to check that $\p_+ \l = \mathrm{a}_+$ for all the non-zero modes, but the zero mode contribution is non-vanishing, yielding
\be
T_{++}^{\text{\tiny{CS}}} = T^{(0)\, \text{\tiny{CS}}}_{++} + J(x^+) \frac{\mu \bar{\mathcal{L}}_0}{8\pi G\ell} (1+P'(x^+)) \label{tcs}
\ee

\be
T^{\text{\tiny{CS}}}_{+-}= T^{\text{\tiny{CS}}}_{-+} = T^{\text{\tiny{CS}}}_{--} =0 \nonumber
\ee
The above structure of the stress tensor agrees with our holographic expression as $k\r 0$. 
Note that our argument is  almost identical to the one of  \cite{Guica:2017lia} for the case of $J\bar T$ - deformed free fermions, since  performing a gauge transformation of the form \eqref{gaugetrsol} on the free fermion action is the same as deforming by $J\bar T$.    As we will see in section \ref{matft}, the above contribution of the current to the stress tensor exactly reproduces the field theory result for the deformed spectrum for $k = 0$.

\bigskip

We now have all the necessary ingredients to write down the holographic one-point functions  in the deformed CFT.  As already explained,  the expectation values of the stress tensor $\tilde T$ and current $\tilde J$ are given in terms of the  expectation values in the original CFT in presence of sources, using \eqref{newvevs}. The expectation value of the current is trivially the same $\tilde J_+ = J(x^+)$. The new stress tensor is given by

\be
\tilde T^a _\a = T^a_\a + (e^a_\a + \mu_\a J^a) \, \mu_b T^b_\b J^\b \label{ttgen}
\ee
where $T^a_\a$ includes both the gravitational and the Chern-Simons contribution.
Plugging in \eqref{expt0} and \eqref{tcs} into the expression above, we find that the  components of $\tilde T_{a \a}$  are

\medskip

\setlength{\jot}{10pt}

\begin{empheq}[box=\widefbox]{align}
\tilde T_{++} = \frac{\L (x^+)}{8\pi G \ell} + \frac{2\pi J^2(x^+)}{ k} + J(x^+)  \frac{ \mu \bar{\mathcal{L}}_0}{8\pi G \ell} (1- \mu J(x^+)) \;, \;\;\;\;\;\;\; \tilde T_{-+} =  0 \nonumber \\
\tilde T_{+-} = - \frac{\bar \L (x^- + P(x^+)) P'(x^+)}{8\pi G \ell}\;, \;\;\;\;\;\;\;\tilde T_{--} = \frac{\bar \L(x^- + P(x^+))}{8\pi G \ell}  \label{ttilde}
\end{empheq}
\setlength{\jot}{2pt}

\medskip

\noindent where, as before, $P'=- \mu J (x^+)$. The contribution from  $\mathcal{L}$  is divergent in the $k \r 0$ limit; this divergence is cancelled by that of the $J^2/k$ term, leaving a finite remainder.

This is our proposed holographic dictionary for the stress tensor. It is easy to check that it satisfies the Ward identity \eqref{newwardid} for $\tilde e^a_\a = \d^a_\a$ and $\tilde{\mathrm{a}}_\a=0$, namely

\be
\p_\l (\tilde T^{a}_\a \d^\l_a ) =0 \label{conseqtt}
\ee
 Note it is crucial that the Chern-Simons contribution only contains the zero mode $\bar{\mathcal{L}}_0$, as otherwise the conservation law would be violated.  The vanishing of the component $\tilde T_{-+} =0$ is precisely what we expect from the $SL(2,\mathbb{R})$ symmetry of the deformed theory. In the following section, we will show that the above values are also in perfect agreement with the one-point functions one obtains from the field theory for $k=0$.

\subsection{The holographic expectation values for $k \neq 0$\label{holovevs}}

The expectation value of the stress tensor $\tilde T_{a\a}$ of the deformed theory when $k\neq 0$ is again given by \eqref{ttgen}, but we now  plug in the expectation values \eqref{expt0} that follow from the standard AdS$_3$/CFT$_2$ dictionary.  Noting that $T_{--} = \frac{1}{\mu} \mathcal{A}$, with $\mathcal{A}$  given in \eqref{solA}, the holographic expectations we thus find are


\medskip

\setlength{\jot}{10pt}

\begin{empheq}[box=\widefbox]{align}
& \hspace{1.9cm} \tilde T_{++} = \frac{\L (x^+)}{8\pi G \ell} + \frac{2\pi \mathcal{J}^2(x^+)}{ k}  \;, \;\;\;\;\;\;\;\;\;\; \tilde T_{-+} =  0 \nonumber \\
&\tilde T_{+-} = - \frac{4\pi}{k \mu^2} \left(1- \sqrt{1- \frac{\mu^2 k \bar{\mathcal{L}}}{16 \pi^2 G \ell}} \right) P'(x^+)\;, \;\;\;\;\;\;\;\tilde T_{--} = \frac{4\pi}{k \mu^2} \left(1- \sqrt{1- \frac{\mu^2 k \bar{\mathcal{L}}}{16 \pi^2 G \ell}} \right) 
  \label{ttildenzk}
\end{empheq}
\setlength{\jot}{2pt}

\noindent  As we already noted, we currently do not understand the exact relation between $\mathcal{J} (x^+)$ and $J(x^+)$, so this dictionary is incomplete.  We do however have a proposal for how to relate their zero modes, which we present in section \ref{matft}.

\section{Checks and predictions \label{checkp}}

In section \ref{jtrev}, we have reviewed how the energy-momentum-charge eigenstates in the $J\bar T$-deformed CFT, labeled by the original conformal dimensions $h_{L,R}$ and the charge $Q_0$, depend on the deformation parameter $\mu$. For $h_{L,R} \gg c$, we expect that these eigenstates\footnote{Or, rather, collections of such eigenstates in a narrow energy band.} are modeled by charged black holes in the bulk. One of the most basic checks  of the holographic dictionary we proposed is to show that the energy of black holes obeying the modified boundary conditions \eqref{afbndc}, computed using the holographic stress tensor of the previous section, agrees with the field-theoretical expectations.

In order to be able to perform this match, we first need to  identify 
the black hole solutions  that correspond to the states labeled by $h_{L,R},  Q_0$ in \eqref{oldspec} or \eqref{moder}.  This can be done by
requiring that their thermodynamic properties (calculated from the horizon area and the  identifications of the euclidean solution) agree with the field-theoretical answers, which we briefly review. 

Since $J\bar T$ induces  a continuous deformation of the spectrum, we do not expect the degeneracy of states to change as a function of $\mu$. Thus, the entropy is the same as the original CFT entropy when expressed in terms of the  variables $h_{L,R}$, $Q_0$

\be
S = 2 \pi \sqrt{\frac{c}{6} \left(h_L - \frac{c}{24} -\frac{Q_0^2}{k}\right)} + 2 \pi \sqrt{\frac{c }{6} \left( h_R - \frac{c}{24}\right) } \label{cftent}
\ee
Replacing $h_{L,R}$ by  their expressions in terms of the left/right energies \eqref{model}, \eqref{specfler}, one can easily derive the thermodynamic properties of the deformed theory. In two dimensions, it is natural to introduce the left/right temperatures $T_{L,R}$, which are conjugate to $E_{L,R}$ 

\be
\d S = \frac{1}{T_L} \d E_L + \frac{1}{T_R} \d E_R \label{firstlaw}
\ee
Using this definition, one finds that the left/right temperatures $\tilde T_{L,R}$ in the $J\bar T$ - deformed CFT are related to their undeformed counterparts via 

\be
\tilde T_L=T_L\;, \;\;\;\;\;\; \frac{1}{\tilde T_R} = \frac{R - \mu Q}{R} \frac{1}{T_R} - \frac{\mu Q}{ R}\frac{1}{T_L} \label{newtemp}
\ee
at fixed $ Q_0$. If $k=0$, we can simply replace $Q$ by $Q_0$ in the expression above. If $k \neq 0$, we note that $Q$  has a piece that depends on the temperatures, so the relation between the temperatures at fixed $Q_0$ is more involved\footnote{Concretely, the relation between the right-moving temperatures becomes 
\be
\frac{1}{\tilde T_R} = \frac{1}{T_R} \sqrt{\left(\frac{R - \mu Q_0}{R} \right)^2 - \frac{\mu^2 \pi^2 c k}{6} \, T_R^2} + \frac{1}{T_L} \left(  \sqrt{\left(\frac{R - \mu Q_0}{R } \right)^2 - \frac{\mu^2 \pi^2 c k}{6}\, T_R^2} -1 \right) \label{reltrknz}
\ee}. 
 Note the above implies that the usual temperature $T_H$ (conjugate to the total energy $E= E_L + E_R$) is given by
\be
\tilde T_H = \frac{T_H}{1 - \mu Q/ R}
\ee
as also found in \cite{Guica:2017lia} for $Q=Q_0$. These computations are valid when $R > \mu Q$.

In the following subsection, we
%
%
%
%
 construct the black hole solutions obeying the modified boundary conditions that correspond to the states labeled by $h_{L,R}$.   To ensure this, we match their parametres to those in the field theory by requiring that their entropy (calculated from the horizon area) matches that of the field theory. For the case of vanishing anomaly ($k=0$), we are able to cross-check this identification by showing that the black hole temperatures (calculated from the  identifications of the euclidean solution) match  the field-theoretical values \eqref{newtemp}. Having fixed the parameters in this way, we show in \ref{matft} that the energy and angular momentum of these black holes also match the field theory expectation, as required by the first law of black hole mechanics. This provides a non-trivial consistency check for our method and the proposed holographic dictionary \eqref{ttilde} at $k=0$. We furthermore show that a similar identification of the parameters  reproduces the correct spectrum also for $k \neq 0$.  

In section \ref{asymm} we use the special properties of the holographic stress tensor \eqref{ttilde} to construct an infinite set of conserved charges in the deformed theory and we compute on the algebra they  satisfy.


\subsection{Black holes and their thermodynamics}

As shown in section \ref{asyjtbar}, gravitational solutions dual to classical states in the deformed theory are obtained by simply applying the coordinate transformation \eqref{coordtr} to the known AdS$_3$ solutions. To obtain a black hole, the seed AdS$_3$ solution is the  BTZ black hole, with metric 

\be
ds^2 = - \frac{(\rho^2-\rho_+^2)(\rho^2-\rho_-^2)}{\rho^2} dt^2 + \frac{\ell^2 \rho^2 d \rho^2}{(\rho^2-\rho_+^2)(\rho^2-\rho_-^2)} + \rho^2 \left(d\hat \varphi + \frac{\rho_+\rho_-}{\rho^2} dt\right)^2 \label{btzmet}
\ee
where $\hat \varphi \sim \hat \varphi + 2 \pi$ and $\rho_\pm$ are the horizon radii. The coordinate $\hat \varphi$ is related to the compact boundary coordinate $\varphi$ via $\varphi= \frac{R}{2\pi} \hat \varphi$. To make the connection with black hole thermodynamics, it is useful to introduce the dimensionless temperatures $\hat T_{L,R} $ via 

\be
\hat T_{L,R} = \frac{\rho_+ \pm \rho_-}{2\pi \ell}= \frac{R}{2\pi}\,T_{L,R} 
\ee
which are the conjugate potentials  to the left/right energies of the black hole defined as in \eqref{firstlaw}.
These temperatures can be easily read off  from the identifications of the euclidean BTZ solution \cite{Maldacena:1998bw}, obtained by the analytic continuation $t \r i \tau$ 

\be
\hat \varphi + i \tau \sim \hat \varphi + i \tau + 2 \pi m + \frac{i n}{\hat T_L} \;, \;\;\;\;\;\; m, n \in \mathbb{Z}
\ee
In euclidean BTZ,  $\rho_-$ is purely imaginary, so $\hat T_{L,R}$ become complex conjugate to each other.   Analytically continuing back to the Lorentzian solution, $\hat \varphi + i \tau \r \hat x^+$ and $\hat \varphi -i \tau \r \hat x^- $, where $\hat x^\pm = \hat \varphi \pm t$ are the rescaled null coordinates. The above identifications of the euclidean coordinates formally act on the Lorenzian null coordinates  as 

\be
\hat x^+ \sim \hat x^+ + 2 \pi m+ \frac{i n}{\hat T_L} \;, \;\;\;\;\;\;\hat x^- \sim \hat x^- + 2 \pi m- \frac{i n}{\hat T_R}
\ee
where $\hat{T}_{L,R}$ have gone back to being real and independent. 

To read off the conserved charges, it is useful to put the metric \eqref{btzmet} in Fefferman-Graham form\footnote{The relationship between $\rho$ and $z$ is $\rho = \frac{1}{z} \sqrt{1+ \frac{z^2}{2} (\rho_+^2+\rho_-^2) + \frac{z^4}{16} (\rho_+^2-\rho_-^2)^2}$; the horizon is at $z = \frac{2}{\sqrt{\rho_+^2-\rho_-^2}}$.}
\be
ds^2 = \frac{\ell^2 dz^2}{z^2} + \frac{4\pi^2 dx^+ dx^-}{z^2 R^2} +16 \pi^2 G \ell \hat h_L \frac{(dx^+)^2}{ R^2}+16 \pi^2 G \ell \hat h_R \frac{(dx^-)^2}{ R^2}  + \frac{z^2}{ R^2} 64 \pi^2 G^2 \ell^2 \hat h_L \hat h_R \, dx^+ dx^- \label{fgbtz}
\ee
where $x^\pm = \frac{R}{2\pi} \hat x^\pm = \varphi \pm t$ and $t$ has been rescaled with respect to the one in the previous equation by a factor of $R/2\pi$. We have defined

\be
\hat h_{L,R}  = \frac{(\rho_+ \pm \rho_-)^2}{16 G \ell} = \frac{\pi^2 c}{6}\, \hat T_{L,R}^2 \;, \;\;\;\;\;\;\; c= \frac{3\ell}{2G}
\ee
Computing the left/right energies by integrating the holographic stress tensor \eqref{holost} over $\varphi$, we find that $ \hat h_{L,R} = E_{L,R} R$, so $\hat h_{L,R}$ are related to the CFT conformal dimensions via $\hat h_{L,R} = h_{L,R} - c/24$
. If the black hole also carries a left-moving $U(1)$ charge $Q_0$, then the relationship between the left-moving energy and $\hat h_L$ is shifted by the Chern-Simons contribution \eqref{tcsnaive} as 

\be
E_L R = 2\pi (\hat h_L + Q_0^2/k) = 2\pi( h_L - c/24) \label{relhhh}
\ee
While the metric \eqref{fgbtz} differs from the usual Fefferman-Graham form by a rescaling of $z$ by $R/2\pi$, this form is extremely useful for tracking the $R$ dependence of the various quantities we read off from the geometry, which will be important later. In particular, one can  easily check, for example, that the line element is completely independent of $R$, as can be seen by rewriting it in terms of the rescaled coordinates $\hat x^\pm$, which are identified mod $2\pi$. This implies, in particular, that the black hole entropy is $R$-independent, and reads 
\be
S = \frac{\mathcal{A}_{horizon}}{4 G} = \frac{\pi \rho_+}{2G} = 2\pi \sqrt{\frac{c\, \hat h_L}{6}} + 2\pi \sqrt{\frac{c \,\hat h_R}{6}}
\ee
which is of course identical to \eqref{cftent}.

%

We would now like to construct black holes in the deformed theory such that their entropy is identical to that of BTZ, when written in terms of $h_{L,R}$. 
Such black holes are obtained by performing the coordinate transformation \eqref{coordtr}  
%
on the undeformed solution. For the energy-momentum-charge eigenstates, $J$ is a constant, $J = Q_0/R$, so the coordinate transformation \eqref{coordtr} simplifies to

\be
x'^+ = x^+ \;, \;\;\;\;\;\; x'^-   = x^- - \frac{ \mu Q_0}{ R} \, x^+\label{trxpx}
\ee
These coordinates have identifications
\be
x'^+ \sim x'^+ + R \;, \;\;\;\;\;\; x'^- \sim x'^- +  R - \mu Q_0 
\ee
We will also find it useful to introduce the rescaled coordinates $\hat{x}'^\pm$

\be
\hat x'^+ = \frac{2\pi x'^+}{R} \;, \;\;\;\;\;\; \hat x'^- = \frac{2\pi x'^-}{R-\mu Q_0}
\ee
which are identified mod $2\pi$. Since  the black hole entropy only depends on $\hat h_{L,R}$, we should require that the deformed metric  be the same as \eqref{fgbtz}, when written  in terms of the coordinates $\hat{x}'^\pm$

\be
ds^2  =   \frac{\ell^2 dz^2}{z^2} + \frac{d\hat x'^+ d\hat x'^-}{z^2 } +4G \ell \hat h_L (d\hat x'^+)^2+4G \ell \hat h_R (d\hat x'^-)^2  + 16 G^2 \ell^2 \hat h_L \hat h_R \,  z^2\, d\hat x'^+ d\hat x'^- \label{defbtz}
\ee
By construction, the above black hole has the same horizon area and thus the same entropy as the original BTZ solution. When rewritten in terms of the $x^\pm$ boundary coordinates \eqref{trxpx} and after a simple rescaling of $z$, this metric is precisely of the form presented in section \ref{asyjtbar}
\bea
ds^2 &= &\frac{\ell^2 dz^2}{z^2} + \frac{dx^+ \left(dx^- -  \frac{\mu Q_0}{R} dx^+\right)}{z^2 } +16 \pi^2 G \ell \hat h_L \frac{(dx^+)^2}{ R^2}+16 \pi^2 G \ell \hat h_R \frac{\left(dx^- -  \frac{\mu Q_0}{R} dx^+\right)^2}{ (R- \mu Q_0)^2}  + \nonumber\\
&& \hspace{2 cm} + \, \frac{(4\pi)^4 z^2 G^2 \ell^2 \hat h_L \hat h_R}{ R^2 (R- \mu Q_0)^2} \, dx^+ \left(dx^- -  \frac{\mu Q_0}{R} dx^+\right) \label{defbtzexpl}
\eea
We would now like to check whether the left/right temperatures that we can read off from the identification of the euclideanized  solution  \eqref{defbtz}  are in agreement with the field theory prediction. Using the fact that,  in terms of the hatted coordinates, the metric is the same as \eqref{fgbtz}, smoothness of the euclidean solution requires that
\be
\hat x'^+ \sim \hat x'^+ + \frac{i n}{\hat T_L} \;, \;\;\;\;\;\; \hat x'^- \sim \hat x'^- - \frac{i n}{\hat T_R}
\ee
This translates into the following identifications of $x^\pm$

\be
x^+ \sim x^+ + \frac{i n R}{2\pi \hat T_L} \;, \;\;\;\;\;\; x^- \sim x^- - \frac{i n (R - \mu Q_0)}{2\pi \hat T_R} + \frac{\mu Q_0}{2\pi} \frac{i n }{\hat T_L} \label{defident}
\ee
which imply that  the left/right temperatures of these black holes are identical to the left/right temperatures \eqref{newtemp} in the deformed theory for $Q=Q_0$. This confirms that, at least for $k=0$, \eqref{defbtz} are the correct black hole solutions, whose energies we should compare with \eqref{oldspec}.

Note that the charge was kept fixed thoughout the above discussion; in particular, we did not consider the electric potential contribution to the first law. Including such a term is possible in principle and it would lead to an additional check on the holographic dictionary: namely, whether the gauge field \eqref{defolds}, in presence of appropriate boundary sources $\tilde{\mathrm{a}}_\a$ proportional to the chemical potential, has vanishing holonomy along the contractible (time) circle of the euclidean black hole geometry. We leave this more complete analysis for future work. 

The above analysis shows that if we identify the parameters $h_{L,R}$ such that the area of the horizon matches, then the match of the temperatures follows from the coordinate redefinitions, at least in the $k=0$ case. It would be interesting to extend this match of the temperatures to $k \neq 0$. In particular, it would be intersting to understand if the new relation 
\eqref{reltrknz} can also have a simple interpretation in terms of coordinate shifts, possibly involving an extra dimension from the point of view of which the gauge field is a Kaluza-Klein gauge field.

\subsection{Match to the field theory spectrum \label{matft}}

We  would now like to use the holographic dictionaries \eqref{ttilde} and \eqref{ttildenzk} to read off the conserved charges associated with the metric \eqref{defbtz} and check whether the energies agree with the field theory expressions \eqref{oldspec} and respectively \eqref{moder}. 
We again perform the analysis separately  for  $k=0$ and $k\neq 0$.

\subsection*{\bf{Chern-Simons level} $k=0$}

The metric \eqref{defbtzexpl} is of the form \eqref{asymet}, with 

\be
\L = \frac{16\pi^2\hat h_L G \ell}{R^2} \;, \;\;\;\;\;\;\;\; \bar \L =\bar \L_0 = \frac{16\pi^2 \hat h_R G \ell}{  (R- \mu Q_0)^2} \;, \;\;\;\;\;\mathcal{J}= J =   \frac{Q_0}{ R}
\ee
Plugging these into the expectation value of the deformed stress tensor \eqref{ttilde}, we find

\be
\tilde T_{++} =\frac{2\pi}{ R} \left(  \frac{\hat h_L }{R} + \frac{Q_0^2}{ k R} +  \frac{ \mu Q_0 \hat h_R}{ R (R-\mu Q_0)}\right)  
\ee

\be
\tilde T_{--} = \frac{2 \pi\hat h_R}{ (R-\mu Q_0)^2} \;,\;\;\; \;\;\;\;\;\; \tilde T_{+-} =  \frac{\mu Q_0}{ R} \cdot \frac{2\pi\hat h_R}{ (R-\mu Q_0)^2}
\ee
and $\tilde T_{-+} =0$, as before. The conserved charges associated to translations along a boundary (Killing) vector $\xi^\a$ are given by 
\be
Q_\xi = \int_{\mathcal{P}}\!\! d \varphi \,  n^a \tilde{T}_{a\a} \xi^\a = \int_0^{ R}\!\!\! d\varphi \, (\tilde T_{+\a} -  \tilde T_{-\a} )\, \xi^\a \label{consch}
\ee
where $\mathcal{P}$ is a constant $t$ slice of the cylinder on which the deformed theory is defined  and $n^a=\p_t$ is the unit vector normal to it. Note that the integral is \emph{not} performed over the induced metric at the boundary of the asymptotically AdS$_3$ spacetime \eqref{modmet}, but rather over an abstract boundary whose metric, previously denoted as $\tilde \g_{\a\b}$, is flat.

The left-moving energy of the black hole is the conserved charge associated with $\xi =\p_+$ 

\be
E_L = \int_0^{ R}\!\!\! d\varphi \, \tilde T_{++}  =\frac{2\pi (h_L- \frac{c}{24})}{ R}+ \frac{\mu Q_0  \cdot 2\pi( h_R - \frac{c}{24})}{ R ( R -\mu Q_0)}
\ee
where we used \eqref{relhhh}. The right-moving energy is the charge associated with $\xi= - \p_-$

\be
E_R = \int_0^{ R}\!\!\! d\varphi \, (\tilde T_{--} -  \tilde T_{+-} )=
\frac{2\pi (h_R-\frac{c}{24})}{R- \mu Q_0}
\ee
Both expressions are in perfect agreement with the field theory result \eqref{oldspec}. This constitutes a rather  non-trivial check of our proposed holographic dictionary \eqref{ttilde}.

\subsection*{\bf{Non-zero Chern-Simons level}}

In the case of a non-zero Chern-Simons level, some of the ingredients of the holographic dictionary for the $J\bar T$ - deformed CFT are  less understood: first, we do not have a first principles derivation of how $\mathcal{J}(x^+)$ and $J(x)$ are related; and second, we only have one input, coming from matching the entropy, for how the black hole parameters should be identified with their CFT counterparts. This should be contrasted to the $k=0$ analysis of the previous subsection, where we matched  both the entropy and the temperatures.
%

We will nevertheless be able to show that, by making  minimal assumptions about what $\mathcal{J}$ and the black hole parameter identification should be in energy-momentum-charge eigenstates, we obtain a perfect match to the field theory spectrum \eqref{moder} for finite $k$. 
%
%
%
%
%
Concretely, the identification we propose between the CFT parameters and the gravitational solution is: 
\be
\mathcal{L} = \frac{16\pi^2 \hat h_L G \ell}{R^2} \;, \;\;\;\;\;\; \bar{\mathcal{L}} = \frac{16\pi^2 \hat h_R G \ell}{(R-\mu Q_0)^2} \;, \;\;\;\;\;\; \mathcal{J} = \frac{Q}{ R} \;, \;\;\;\;\;\; J = \frac{Q_0}{R}\label{lev}
\ee
where $Q$ is given in \eqref{newq}. We have taken the values of $\mathcal{L}$ , $\bar{\mathcal{L}}$ and $J$ to be the same as in the previous section, which means that we keep intact the shift \eqref{defolds} of the sources by the CFT chiral current. The identification of $\mathcal{J}$ with $Q$ can be justified as follows: in the bulk, we have a chiral current $\mathcal{J} (x^+)$ in presence of an external gauge field $\mathrm{a}_\a = \mu_a T^a{}_\a$, which satisfies $(\g^{\a\b} - \e^{\a\b}) \mathrm{a}_\b =0$.  Following our argument in section \ref{jtrev}, in presence of the chiral anomaly, the current whose charge is constant along the flow is 

\be
\hat J^\a =\mathcal{J}^\a + \frac{k}{8\pi} (\g^{\a\b} + \e^{\a\b}) \mathrm{a}_\b = \mathcal{J}^\a + \frac{k}{4\pi}\,  \mu_a T^{a\a} \label{shiftchmu}
\ee
or, in components 

\be
\hat J^+ =  \frac{\mu k}{2 \pi} T_{--} \;, \;\;\;\;\;\; \hat J^- = 2 \mathcal{J} - \frac{\mu k}{2 \pi} P' T_{--}
\ee
%
%
where the indices have been raised with the metric \eqref{modmet}.  The associated conserved charge is

\be
\hat Q = \int_0^R d \varphi \, (\hat J_+ - \hat J_-) = R \mathcal{J} - \frac{\mu k}{4\pi}  \int_0^R d \varphi (P'+1) T_{--}  = R \mathcal{J} - \frac{\mu k}{4\pi}  E_R \label{hqholo}
\ee
where the index on $\hat J^\a$ is now lowered\footnote{ The justification for this change of metric is that it is the contravariant $\hat J^\a$ that should be identified between the two descriptions. } with $\tilde \g_{\a\b} = \eta_{\a\b}$. Since, as we argued before, we expect the charge associated with $\hat J$ to be constant along the flow,  $\hat Q = Q_0$, we deduce that the charge associated with $\mathcal{J}$ should be the instantaneous charge  $Q$. It is interesting to note that while our analysis in \ref{jtrev} was performed infinitesimally in $\mu$, the shift \eqref{hqholo} of the holographic charge appears to work in the same way at finite $\mu$. 

Once the assignments \eqref{lev} are accepted, it is straightforward to read off the modified spectrum from holography. 
%
%
%
%
Using \eqref{ttildenzk}, the expression for the left-moving energy is 

\be
E_L = \int_0^{ R} \!\!\!d\varphi \, \tilde{T}_{++} = 2\pi \left(\frac{ h_L- \frac{c}{24}}{R}-\frac{Q^2_0}{k R} + \frac{Q^2}{k R} \right)
\ee
which  matches \eqref{model}. The right-moving energy is given by

\be
E_R =  \int_0^{ R} \!\!\!d\varphi \, (\tilde{T}_{--} -\tilde{T}_{+-}) =  R \left(1- \frac{\mu Q_0}{ R}\right) \tilde{T}_{--}
\ee
where the expression for $\tilde T_{--}$ is given in \eqref{ttildenzk}. Plugging in the value of $\bar{\mathcal{L}}$, we obtain

\be
E_R = \frac{4\pi}{k \mu^2} ( R -  \mu Q_0 ) \left(1-\sqrt{1- \frac{ \mu^2 k ( h_R- \frac{c}{24})}{( R-\mu Q_0)^2}}\right)
\ee
which  precisely matches \eqref{moder}.

The fact that we can match the field theory spectrum for $k\neq 0$ with a very small number of assumptions is encouraging. In particular, the only missing piece is a better understanding of how the chiral currents are related in the deformed versus the undeformed CFT, which is a field-theory, rather than a holography, question. 
 It would also be nice to better understand why the shift \eqref{shiftchmu} works so well for $\mu$ finite. When the gauge field is realized geometrically as a Kaluza-Klein gauge field, it would be interesting to understand whether a  coordinate transformation can explain the source-vev mixing that we found. 

\subsection{Symmetry enhancement \label{asymm}}

An interesting question is whether the global $SL(2,\mathbb{R})_L \times U(1)_R \times U(1)_J$ symmetries of the $J\bar T$ - deformed CFT allow for an infinite-dimensional extension. For the case of \emph{local} two-dimensional QFTs with an $SL(2,\mathbb{R})_L  \times U(1)_R$ global symmetry, this question has been previously studied in \cite{Hofman:2011zj}, who showed that the $SL(2,\mathbb{R})_L$ symmetry is enhanced, as expected, to a left-moving Virasoro symmetry, while the $U(1)_R$ is enhanced to either a left-moving Ka\v{c}-Moody symmetry, or to a right-moving Virasoro. 

The case of $J\bar T$-deformed CFTs is somewhat different, because the  theory is non-local along the $U(1)_R$, a.k.a. $x^-$, direction. Thus, we do not necessarily expect to find a local infinite-dimensional symmetry. 
To study the symmetry enhancement, there are two possible approaches: either to use the special properties of the stress tensor in the deformed theory to construct an infinite set of conserved charges, or to study the asymptotic symmetries of the dual spacetime. In the following, we will use both methods to argue that the above global symmetries are enhanced to infinite-dimensional ones, the  $SL(2,\mathbb{R})_L \times U(1)_J$ to a  left-moving Virasoro-Ka\v{c}-Moody symmetry as before, and the $U(1)_R$ to a state-dependent, effectively non-local Virasoro symmetry. 

\subsubsection*{i) Direct construction of the conserved charges}

The  holographic one-point functions \eqref{ttilde} show that the stress tensor of the deformed CFT obeys

\be
\tilde{T}_{-+} =0 \;, \;\;\;\;\;\; \tilde{T}_{+-} = - P'\, \tilde{T}_{--}  =   \mu J \, \tilde{T}_{--} 
\ee
for arbitrary CFT states with a classical bulk dual. The first equation simply follows from the $SL(2,\mathbb{R})_L$ symmetry of the theory and holds also as an operator equation; the second is derived from holography, though there may also exist a purely CFT derivation, valid for more general states. Together with the conservation equations \eqref{conseqtt}, they imply the following spacetime dependence of the stress tensor components

\be
\tilde T_{++} = \tilde T_{++} (x^+)\;\;\;\;\;\; \tilde{T}_{--} =  \tilde{T}_{--} (x^- + P(x^+)) \label{relstc}
\ee
Using the above, it is easy to construct an infinite family of conserved charges. These are given by

\be
Q_{\chi_L} = \int_0^{ R} d \varphi \, \tilde{T}_{++}\, \chi_L (x^+) \;, \;\;\;\;\;\;\;\;\;Q_{\chi_R} = \int_0^{ R} d \varphi \, \left(\tilde{T}_{--}- \tilde{T}_{+-} \right) \chi_R (x^-+P(x^+)) \label{qdiffs}
\ee
where $\chi_{L,R}$ are arbitrary functions of their respective arguments, and each of their Fourier modes is associated with a separate conserved charge. These charges are conserved due to conservation of the currents $ \tilde T_{\a\b} \,\xi_{L,R}^\b$ for $\xi_L = \chi_L (x^+) \p_+$ and $\xi_R = -\chi_R (x^-+P(x^+))\p_-$, respectively, where the derivative is computed with respect to the flat  ``boundary'' metric $\tilde \g_{\a\b} =\eta_{\a\b}$ of the deformed CFT. Note that while $\xi_L$ is a conformal Killing vector of the ``boundary'' metric $\eta_{a\b}$, as is usually the case, $\chi_R$ is not; rather, its form is dictated by charge conservation, together with the relation \eqref{relstc} between the stress tensor components.  Using it, the formula for the right-moving conserved  charges reduces to

\be
Q_{\xi_R} = \int_0^{ R} d \varphi \,(1-  \mu J(x^+)) \,\tilde{T}_{--}(x^-+P(x^+)) \, \chi_R (x^-+P(x^+)) \label{qxir}
\ee
whose integrand is a total $\varphi $ derivative.
Note this is very similar to the formula for the right-moving Virasoro generators in a two-dimensional CFT, except that the argument has been shifted by a state-dependent function of $x^+$.  This implies that the action of these symmetry generators on local operators is effectively non-local. 

In addition to the above translational symmetries, there is an infinite enhancement of the $U(1)_J$ global symmetries, generated by 
\be
Q_\lambda = \int_0^{ R} d \varphi \, \l(x^+) \mathcal{J}(x^+) \label{qgaug}
\ee
for an arbitrary function $\l (x^+)$, as follows from the conservation of the current $\l(x^+) \, \mathcal{J}^\a$. 

Thus, we have shown that each symmetry factor acquires an infinite-dimensional extension, in a way very similar to what happens in two-dimensional CFTs. Two of these enhanced symmetries, $Q_{\chi_L}$ and $Q_\lambda$, are local, while the remaining one, $Q_{\chi_R}$ is state-dependent and appears non-local. One natural question is to find the algebra satisfied by these conserved charges, which before the deformation is Virasoro$_L \times$ Virasoro$_R \times U(1)$ Ka\v{c}-Moody. In the following, we will use the complementary approach of asymptotic symmetries to argue that the algebra stays the same, even though the generators \eqref{qxir} have been deformed.

\subsubsection*{ii) Asymptotic symmetries}

The enhanced symmetries of the field theory can be often obtained  from an analysis of the asymptotic symmetries of the dual spacetime. These are the   diffeomorphisms and gauge transformations that preserve the boundary conditions on the metric and gauge field, while leading to non-trivial conserved charges. 
An important result that we will be using in this section is the representation theorem \cite{Barnich:2007bf}, which states that the Dirac bracket algebra of the conserved charges  is isomorphic to the (modified) Lie bracket algebra $\{\,\,,\,\}_*$ of the associated asymptotic symmetry generators, up to a possible central extension

\be
\{ Q_{\mathrm{g}_1},Q_{\mathrm{g}_2} \}_{D.B.}  =  Q_{\{\mathrm{g}_1,\mathrm{g}_2\}_{*}} + \mathcal{K}_{\mathrm{g}_1,\mathrm{g}_2} [\bar{\Phi}]
\ee
that only depends on the reference background $\bar{\Phi}$. 
This means that  we do not need to explicitly compute  the charges 
to find the algebra they satisfy, but we  can simply  infer it from the algebra of the asymptotic symmetries, under the assumption that the charges will turn out to be integrable. 

In our case, each asymptotic symmetry is implemented by a pair $\mathrm{g} = (\xi, \Lambda)$ of a diffeomorphism and a gauge transformation. Additionally, both symmetry parameters can depend on the background fields $\Phi = \{\mathcal{L}, \bar{\mathcal{L}}, J \}$ that parametrize the asymptotic solution: $\xi = \xi [\chi, \Phi], \; \Lambda = \Lambda [\chi,\Phi]$, where $\chi$ denotes the  parameter of the transformation.  In presence of symmetries simultaneously associated with diffeomorphisms and gauge transformations, the algebra of the asymptotic symmetries is \cite{Compere:2007az}

\be
\{(\xi_1,\Lambda_1),(\xi_2,\Lambda_2)\}_{*}  = \bigl(\{\xi_1,\xi_2\}_*,\{\Lambda_1,\Lambda_2\}_*\bigr)
\ee
where $\{\xi_1,\xi_2\}_*$ is given by a modified Lie bracket \cite{Barnich:2010eb} (see also \cite{Compere:2015knw})

\be
\{\xi_1,\xi_2\}_* = [\xi_1,\xi_2]_{L.B.} - \left(\d_{\chi_1}\! \Phi \; \p_\Phi \xi_2 - \d_{\chi_2} \!\Phi \; \p_\Phi \xi_1 \right) \label{modlieb}
\ee
which takes into account the field-dependence of the diffeomorphisms. 
The bracket of the gauge transformations is given by
\be
\{\Lambda_1,\Lambda_2\}_* = \mathcal{L}_{\xi_1} \Lambda_2 - \mathcal{L}_{\xi_2} \Lambda_1- \left(\d_{\chi_1}\! \Phi \; \p_\Phi \Lambda_2 - \d_{\chi_2}\! \Phi \; \p_\Phi \Lambda_1 \right) \label{gaugeb}
\ee
where we subtracted a similar correction due to field-dependence of the gauge parameter  from the bracket quoted in \cite{Compere:2007az}. 

The asymptotic form of the metric is left invariant by a combination of diffeomorphisms and a gauge transformation that i) preserve the radial gauge for the metric and the gauge field and ii) leave invariant the asymptotic relations \eqref{afbndc} between sources and expectation values. They are parametrized by the same three functions $\chi_L (x^+)$, $\chi_R (x^- + P(x^+))$ and $\lambda(x^+)$ that we introduced above. 
 The diffeomorphism component of the asymptotic symmetries is

\be
\xi_L = \chi_L \, \p_+ - \frac{z^2  \ell^2}{2}\,  \chi_L'' \p_- + \frac{z}{2}\, \chi_L' \,\p_z + \ldots \nonumber
\ee

\be
\xi_R = \left( \chi_R + \frac{z^2 \ell^2}{2}\, P'(x^+) \,\chi_R'' \right) \p_- - \frac{z^2 \ell^2}{2} \chi_R'' \,\p_+ + \frac{ z}{2} \chi_R'\, \p_z  + \ldots \label{as}
\ee

\be
\xi_\Lambda = - \frac{\mu k }{4\pi} \l(x^+)\, \p_- \nonumber
\ee
The first two diffeomorphisms are the direct analogues of Brown-Henneaux diffeomorphisms in AdS$_3$; the last one is the same as the generator of Ka\v{c}-Moody symmetries of \cite{Compere:2013bya}. Note that  $\xi_R$ is now a conformal Killing vector of the boundary metric \eqref{modmet}. 

The accompanying gauge transformations take the form

\be
\Lambda_L = \frac{\mu\ell \bar{ \mathcal{L}}}{16\pi G}  \,\chi_L'' z^2 + \ldots 
\ee

\be
 \Lambda_R =  \int \!\!\mathcal{A} \, \chi'_R - \left(\frac{ \mu \ell}{16\pi G } - \frac{2\pi\ell^2}{k} z^2 J(x^+)\right)  \chi_R''+\ldots
\ee

\be
\Lambda_\Lambda = \lambda (x^+)
\ee
%
%
where $\mathcal{A}$ has been defined in \eqref{formgf}. 
The $\ldots$ denote terms proportional to higher powers of $z$, which are not expected to contribute to the conserved charges.  

  The combined action of the diffeomorphisms and the gauge transformation on the functions that specify the solution is given by
\be
\d \mathcal{L} =  2 \mathcal{L}\, \chi_L'+\mathcal{L}'\chi_L - \frac{\ell^2}{2} \chi_L''' \;, \;\;\;\;\;\;  \d \bar{\mathcal{L}} =  2 \bar{\mathcal{L}}\, \chi_R'+ \bar{\mathcal{L}}'\, \chi_R - \frac{\ell^2}{2} \chi_R'''\label{delgvv}\ee

\be
  \d \mathcal{J}(x^+) = \p_+ \left(\chi_L \mathcal{J}(x^+) + \frac{k}{4\pi}\lambda (x^+)\right)
\ee
which is identical to the transformation of the analogous functions specifying a general asymptotically AdS$_3$ solution (with Dirichlet boundary conditions) under the Brown-Henneaux diffeomorphisms and a holomorphic gauge transformation. However, due to the dependence of its argument on $J$, $\bar{\mathcal{L}}$ additionally transforms under the left-moving symmetries parametrized by $\chi_L$ and $\l$ as

\be
\d \bar{\mathcal{L}} = - \mu \left(\chi_L J(x^+) + \frac{k}{4\pi}\lambda (x^+)\right) \bar{\mathcal{L}}'
\ee
The above equations provide all the variations $\d_\chi \Phi$ needed to compute the algebra of the asymptotic symmetries. Denoting $ (\xi_L, \Lambda_L)= \mathrm{g}_L$ etc., we find

\be
\left\{\mathrm{g}_L [\chi_L], \mathrm{g}_L [\eta_L]\right\}_* = \mathrm{g}_L  [\chi_L \eta'_L -  \chi'_L\eta_L] \;, \;\;\;\;\;\;\{\mathrm{g}_L[\chi_L],\mathrm{g}_\Lambda[\lambda] \}_*=\mathrm{g}_\Lambda[\chi_L \lambda'] 
\ee

\be
\left\{\mathrm{g}_R [\chi_R], \mathrm{g}_R [\eta_R]\right\}_* = \mathrm{g}_R  [\chi_R \eta'_R -  \chi'_R\eta_R]
\ee

\be
\{\mathrm{g}_L[\chi_L],\mathrm{g}_R[\chi_R] \}_* =  \{\mathrm{g}_\Lambda[\lambda],\mathrm{g}_R[\chi_R] \}_*=  \{\mathrm{g}_\Lambda[\lambda],\mathrm{g}_\Lambda[\l'] \}_* =0
\ee
The use of the modified brackets \eqref{modlieb} - \eqref{gaugeb} is essential to show that the various cross-commutators between the symmetries vanish. To find the asymptotic symmetry algebra, we consider as usual the individual Fourier modes of the above gauge transformations, i.e. we take

\be
\chi_L (x^+)  =   \frac{R}{2\pi} \, \exp \left({\frac{2\pi i m x^+}{R}}\right) \equiv \chi_m\;, \;\;\qquad\;\;\;\;\;\; \l (x^+) = \frac{R}{2\pi}  \, \exp \left({\frac{2\pi i m x^+}{R}}\right)\equiv \l_m  \nonumber
\ee

\be
 \chi_R (x^- + P(x^+))  =  \frac{R-\mu Q_0}{2\pi} \, \exp \left({\frac{2\pi i m (x^- + P(x^+))}{R-\mu Q_0}} \right) \equiv\tilde \chi_m
\ee
with $m \in \mathbb{Z}$. The non-zero commutators are

\be
\{\chi_m, \chi_n \}_* = - i \, (m-n) \chi_{m+n}  \;, \;\;\;\;\;\; \{\chi_m,\l_n\}_* = i \, n \l_{m+n} \;, \;\;\;\;\; \{\tilde\chi_m, \tilde\chi_n \}_* =- i \, (m-n) \tilde\chi_{m+n}
\ee
which correspond to two copies of the Witt algebra and one copy of $U(1)$  Ka\v{c}-Moody. Note that in order to make sense of the $\tilde \chi_m$ commutators, we work in a sector of fixed total charge $Q_0$.

The modified Lie bracket of the asymptotic symmetry generators only yields the algebra of the corresponding conserved charges 
up to a possible central extension $\mathcal{K}_{g_1,g_2}$ - a term that only depends on the background that cannot be absorbed into a redefinition of the conserved charges. The simplest way to compute it is from the definition of the commutator

\be
\{ Q_{\mathrm{g}_1},Q_{\mathrm{g}_2} \}= \d_{\mathrm{g}_2} Q_{\mathrm{g}_1}
\ee
together with the expressions  \eqref{qdiffs}, \eqref{qgaug} for the charges and the variation \eqref{delgvv} of the background fields. Since we are only interested in the central extensions,  we only need to keep the inhomogenous term in the symmetry variations \eqref{delgvv}. These terms are identical to their AdS$_3$ counterparts, which implies that the central extensions are the same as before the deformation. The only potentially non-trivial additional contribution we find comes from the term linear in $J$ in the expression for $\tilde T_{++}$ in \eqref{ttilde}, which would imply a central extension of the  commutator between the left-moving Virasoro generators and the Ka\v{c}-Moody ones; however, the $n$ dependence of this term is such that it can be absorbed into a redefinition of the current. Consequently, the algebra of the symmetry generators that we find is identical, including all the central extensions, to that of a CFT with a left-moving $U(1)$ current; the only difference  is that the spacetime dependence of the right-moving Virasoro generators becomes  state-dependent.

\section{Discussion}

In this article, we have worked out the holographic interpretation of $J\bar T$ deformed CFTs. Even though the deforming operator is irrelevant, we have shown that the usual treatment of the double-trace deformation in terms of mixed boundary conditions  for the bulk fields yields results that are in perfect agreement with the ones previously derived from field theory, at least for the case of vanishing chiral anomaly. 

Since the variational principle we employed here should be equivalent, in principle, with the direct evaluation of the deformed generating functional at large $N$, we expect that our results on  relating sources and expectation values before and after the deformation will  be valid also at a purely field-theoretical level. This  should allow us to compute arbitrary correlation functions in the deformed theory in terms of correlation functions in the original CFT. 
The results can then be compared with the correlators computed using conformal perturbation theory, as well as  with predictions from the holographic dictionary.

There are several points of our analysis that need to be further addressed. First, we would need a better understanding of how the chiral current $J$ is defined along the flow, and how it relates to the undeformed CFT current. In particular, it would be worth confirming, e.g. via a  concrete example, whether the intuitive physical picture we put forth in section \ref{jtrev} is correct. Translating the result of this analysis into holography should allow us to complete the holographic dictionary of section \ref{holovevs}.
 A related  point, which also necessitates an understanding of how the current shifts due to the anomaly, is to reproduce the field-theoretical temperatures  and chemical potential for the charge from the identifications of the euclidean solution  for
  $k\neq 0$, when the relation between the CFT and the deformed temperatures is significantly more involved \eqref{reltrknz}.  
 


One of the main effects of the chiral anomaly on the spectrum of $J\bar T$-deformed CFTs is the appearance of an upper bound on the allowed right-moving energies \cite{Chakraborty:2018vja}. As we saw, in our setup its bulk signature  is the fact that the Chern-Simons gauge field becomes imaginary when  this bound is violated. It would be very interesting to have a more geometric picture of this bound, obtained e.g. by treating the Chern-Simons gauge field as a Kaluza-Klein gauge field associated to an extra dimension, which may allow us to gain a better physical understanding of the significance of this bound.

An interesting outcome of our analysis is that the double-trace $J\bar T$ deformation, rather than completely breaking the right-moving Virasoro symmetries - as one would naively expect from the fact that it is irrelevant on the right - merely deforms it into a non-local version of the Virasoro algebra. 
In principle, it should be possible to derive its presence  from a purely field-theoretical perspective, e.g. by using conformal perturbation theory. It would be interesting to better understand  the  significance of these ``state-dependent'' symmetries, their representations, and, provided they represent proper symmetries, how the states of the $J\bar T$ -deformed theory organise themselves into these representations. 
  A related question is whether the relation \eqref{relstc} between the stress tensor components can be understood at an operatorial level as a consequence of a similarly state-dependent \emph{global}  $SL(2,\mathbb{R})_R$ symmetry, which is then enhanced to an infinite-dimensional symmetry in the usual way\footnote{We thank Diego Hofman for emphasizing this point.}. Of course, the precise definition and meaning of such a global symmetry remains to be understood. 

As we have already noted, the boundary conditions we have derived are very similar to the alternate boundary conditions for AdS$_3$ proposed by \cite{Compere:2013bya}. In that work, since there was no gauge field to compensate for the variation of the boundary metric, the second Virasoro symmetry was absent, and the asymptotic symmetry group consisted solely of the left-moving Virasoro and  Ka\v{c}-Moody algebras. The level of the $U(1)$ Ka\v{c}-Moody algebra was found
to depend on the $\bar{\L}_0$  eigenvalue, and was negative for black hole spacetimes; this fact was interpreted as
corresponding to ergosphere formation.  It would be interesting to embed these boundary conditions in   our setup and understand the interpretation of these ``warped conformal'' symmetries and  state-dependent level  from the perspective of the $J\bar T$-deformed CFT. In particular, it would be interesting if our setup could be used to construct an explicit holographic warped CFT.   


Finally, note that our  method based on the variational principle should straightforwardly apply also to more general deformations of the Smirnov-Zamolodchikov type, such as the $T\bar T$ deformation. Heuristically, it is easy to see that the mixed boundary conditions associated with this deformation will fix a combination of the boundary metric and the stress tensor that,  for the appropriate sign of $\mu$, 
%
 corresponds to fixing the metric on a particular radial slice, 
in agreement with the proposal of \cite{McGough:2016lol}. It would be interesting to study the relation between the expression for the large $N$ generating functional that would be obtained using the variational approach and the recent proposals of \cite{Dubovsky:2017cnj,Cardy:2018sdv}, as well as the conformal perturbation theory results of \cite{Kraus:2018xrn}. 

\subsubsection*{Acknowledgements}

The authors  are grateful to Camille Aron, Alejandra Castro, Geoffrey Comp\`ere,  Diego Hofman, David Kutasov,  Finn Larsen, Ioannis Papadimitriou, Herman Verlinde and Konstatin Zarembo for interesting conversations and  Geoffrey Comp\`ere and Andrew Strominger for comments on the draft. M.G. would like to thank 
the Tsinghua Sanya International Mathematics Forum  for its hospitality while this work was being completed. The work of 
M.G.  was supported by the ERC Starting Grant 679278 Emergent-BH, the Knut and Alice Wallenberg Foundation under grant 113410212 (as Wallenberg Academy Fellow) and the Swedish Research Council grant number 2015-05333. The work of A.B. is supported
by the ANR grant Black-dS-String ANR-16-CE31-0004-01 and the CEA Enhanced Eurotalents Fellowship.

\appendix

\numberwithin{equation}{section}

\section{Current-current deformations \label{cc}}

In this appendix, we study in detail a very simple analogue of the $J\bar T$ deformation, which is a marginal $J\bar J$-type deformation. The main goal is to gain some intuition into the $\mu$-dependence of the $U(1)$ charge due to the chiral anomaly, from both a field-theoretical and a holographic perspective.  In specific calculations, we will  model the currents by free chiral bosons.  

More precisely, the deformation we consider is given by

\be
\frac{\p}{\p \mu} S (\mu) =  - \int d^2 z \, \sqrt{\g} \, (J_1)^\a (\bar J_2)_\a= - \int d^2 z \, J_1 \bar J_2 \label{defjjbar}
\ee
where $J_1$ is a chiral current, $\bar J_2$ is an antichiral one, and the signature is euclidean (the sign of $\mu$ would be opposite in Lorentzian signature). We assome that the chiral anomaly coefficient is $k$ for both currents and that it stays constant along the flow.

\subsection{The spectrum}

The charges associated with the two chiral currents are

\be
Q_1 = \int_0^R \!\! d\varphi \, J_1 \;, \;\;\;\;\; Q_2 = - \int_0^R \!\! d\varphi \, \bar J_2
\ee
The change in the action above corresponds to a change in the  energy levels of the system placed on a circle of circumference $R$ of the form

\be
\frac{\p}{\p \mu} E = - 2 R \langle J_1 \rangle \langle \bar J_2 \rangle = \frac{2 Q_1 Q_2}{R} \label{pejj}
\ee
This would be a very boring deformation if the charges were constant with respect to $\mu$. However, the charges are not constant if we take into account the chiral anomaly. As in section \ref{jtrev}, we can infinitesimally view the $J_1 \bar J_2$ coupling
 as a  gauge field $\mathrm{a}_{\bar z} = - \d \mu \bar J_2$ that couples to the chiral current $J_1$
.
  Due to the anomaly, the current whose charge is conserved along the flow picks up an antichiral component

\be
(\d_\mu  \hat J_1)_{\bar z} =  \frac{ k}{4\pi} \mathrm{a}_{\bar z} =  - \frac{k}{4\pi} \d\mu \, \bar J_2 \label{delj1}
\ee
Since $\mathrm{a}_{\bar z} \propto \bar J_2$ is antiholomorphic, the chiral and antichiral components of $\hat J_1$ are separately conserved. However, the charge that is conserved along the flow is associated to the full non-chiral current $\hat J_1$, as we will be able to see explicitly in section \ref{frbos}. If we insist to just work with the chiral component of $\hat J_1$ (which at infinitesimal level equals $J_1$), the change in its associated charge satisfies


\be
\d_\mu \hat Q_1 = \int_0^R d\varphi \, (\d_\mu \hat J^1_z - \d_\mu \hat J^1_{\bar z} ) = \d_\mu Q_1 - \frac{k}{4\pi} \d \mu\,  Q_2 =0
\ee
which implies that 
\be
\frac{\p}{\p \mu} Q_1 = \frac{k}{4\pi} Q_2 \label{pqjjw}
\ee
Repeating the analysis for $\bar J_2$, we obtain an identical equation with $Q_2 \leftrightarrow Q_1$. The solution is simply

\be
Q_1 = \hat Q_1 \cosh \tilde \mu +\hat Q_2 \sinh \tilde \mu \;, \;\;\;\;\; Q_2 = \hat Q_2 \cosh\tilde \mu + \hat Q_1 \sinh \tilde\mu \;, \;\;\;\;\;\tilde \mu = \frac{\mu k}{4\pi} \label{flowqjj}
\ee
where $\hat Q_{1,2}$ are the conserved charges along the flow. 
The spectrum is determined by the coupled differential equations \eqref{pejj} and \eqref{pqjjw}. Using these equations, it is easy to show that 

\be
E R - \frac{2\pi}{k} (Q_1^2 + Q_2^2) = const \label{jjsf}
\ee
This formula can be written more suggestively as a set of spectral flow equations, using the fact that $P= E_L - E_R$  and $Q_1^2 - Q_2^2$ are constant along the flow

\be
R E_L - \frac{2\pi Q_1^2}{k} = const  \;, \;\;\;\;\;\;\;\;\;R E_R - \frac{2\pi Q_2^2}{k} = const
\ee

\subsection{Free bosons example \label{frbos}}

We would now like to ascertain these general considerations in a simple example. 
We model the currents $J_1, \bar J_2$ as two free bosons and we deform by the $J_1 \bar J_2$ term above\footnote{This is a finite deformation by $J_1\bar J_2$, rather than an instantaneous one; however, since the only effect of the $J\bar J$ flow is to rescale the currents, we have $\l = \l(\mu)$, where $\mu$ is the instantaneous deformation parameter defined in \eqref{defjjbar}.  }. The (Lorentzian) action reads

\be
S = -\frac{1}{2\pi} \int d^2 z \left(\p \phi_1 \bar \p \phi_1 + \p \phi_2 \bar \p \phi_2 - 2\l \p \phi_1 \bar \p \phi_2 \right)
\ee
where $\p, \bar \p = \frac{1}{2}(\p_\varphi \pm \p_t)$, $z, \bar z = \varphi \pm t$.
The equations of motion are still $\Box\phi_1 = \Box \phi_2 =0$, which means that the non-zero modes are either chiral (for $\phi_1$) or antichiral (for $\phi_2$). The normalization has been chosen such that for $\l=0$, each boson has\footnote{In this section only, the normalization of the $J(z)J(0)$ OPE is $k/2z^2$.} $k=1$.  The conserved momenta are

\be
\pi_1 = \frac{1}{2\pi} (\dot \phi_1 +2 \l \bar \p \phi_2 ) 
\;, \;\;\;\;\; \pi_2 = \frac{1}{2\pi} (\dot \phi_2 - 2\l \p \phi_1 )
\ee
and are quantized in units of $1/R$, where $R$ is the circumference of the $\varphi$ circle: $\pi_{1,2} = n_{1,2}/R$, where $n_{1,2} \in \mathbb{Z}$.  
For compact bosons of unit radius, we denote the (integer) winding numbers by $w_{1,2}$, with $ \p_\varphi \phi_{1,2} = 2 \pi  w_{1,2}/R$. Solving for the velocities, we find

\be
\dot \phi_1 = 2\pi \cdot\frac{n_1 - \l w_2+ \l n_2 + \l^2 w_1}{(1-\l^2) R} \;, \;\;\;\;\;\;\dot \phi_2 = 2\pi\cdot\frac{n_2+ \l w_1 + \l n_1-\l^2 w_2}{(1-\l^2) R}
\ee
Taking the intitial current to be (anti)chiral, we have $w_1=n_1$ and $w_2 = - n_2$. 
The canonical commutators read 
\be
\{\phi_1, \dot  \phi_1 - \l \dot  \phi_2 \} = \{\phi_2,\dot   \phi_2 - \l\dot  \phi_1 \} = 2\pi\d(x-x') \;, \;\;\;\;\; \{\phi_2, \dot  \phi_1 - \l  \phi_2 \} = \{\phi_1, \dot  \phi_2 - \l\dot \phi_1 \} =0
\ee
which implies that 

\be
\{\phi_1, \dot \phi_1 \} = \{\phi_2, \dot \phi_2 \} = 2\pi \frac{1}{1-\l^2} \d(x-x') \;, \;\;\;\;\; \{\phi_1, \dot \phi_2 \} = 2\pi\frac{\l}{1-\l^2} \d(x-x') \label{commut}
\ee
This means that if we want the two-point function of the chiral scalars to be canonically normalized, we should consider the rescaled bosons $\tilde \phi_i = \phi_i \, \sqrt{1-\l^2}$. 

There are several conserved currents one can discuss. The current associated with shifts in $\phi_1$ is 

\be
\hat J'^1_z =  \p \phi_1 \;, \;\;\;\;\;\;\hat J'^1_{\bar z} =\bar \p \phi_1 -2 \l \bar \p \phi_2  
\ee
The associated conserved charge

\be
\hat Q'_1 = R ( \p \phi_1-\bar \p \phi_1 + 2\l \bar \p \phi_2) = 2\pi R \, \pi_1 = 2\pi n_1
\ee
does not change with $\l$. Naively, the above current should be  the analogue of the current $\hat J_1$ we considered in the previous subsection, which is not chiral for $\l \neq 0$ and its associated charge is constant along the flow. However, a closer inspection,  e.g. for $\l$ infinitesimal where its relation to $\mu$ in the previous section is simply $\l \sim \mu/(4\pi)$, reveals that the prediction for $\hat J^1_{\bar z}$ from \eqref{delj1} is rather $-\l \bar \p \phi_2$. Indeed, it is easy to see that the combination $\bar \p \phi_1 - \l \p \bar \phi_2 = \pi (w_1-n_1)/R$ is constant along the flow, and in particular it vanishes for an initially chiral current. Thus, the current

\be
\hat J^1_z =  \p \phi_1 \;, \;\;\;\;\;\;\hat J^1_{\bar z} =- \l \bar \p \phi_2  \label{j1}
\ee
is an equally good conserved current, whose charge stays constant along the flow and moreover it is exactly chiral at $\l=0$, including the zero modes. We would thus like to identify $\hat J_1$ with \eqref{j1}.

 From the equations of motion we see that the left and right-moving components of the current are separately conserved, so we are also free to consider the chiral current

\be
J_{1} = \p \tilde \phi_1 = \frac{1}{2} \sqrt{1-\l^2} (\dot \phi_1 + 2\pi w_1/R) 
\ee
whose normalization has been chosen such that the $J_1J_1$ OPE has $k=1$ for all $\l$, as follows from \eqref{commut}.  This current is also conserved, but the associated charge does vary with $\l$

\be
Q_1 =\pi \cdot \frac{n_1 + \l n_2+w_1 - \l w_2}{\sqrt{1-\l^2}}
\ee
Similar comments hold for $\bar J_2$, for which the charge is

\be
Q_2 = \pi \cdot \frac{n_2 + \l n_1- w_2 + \l w_1}{\sqrt{1-\l^2}}
\ee
Setting $\l = \tanh \tilde \mu$,  we note that these are exactly the flow equations \eqref{pqjjw} for the chiral charges as a function of $\mu$. In particular, 
it is easy to check that $Q_1^2 - Q_2^2$ is independent of $\l$

\be
Q_1^2 - Q_2^2 = \pi^2 (n_1^2-n_2^2 + w_1^2-w_2^2 +  2 n_1 w_1 + 2 n_2 w_2)
\ee
The Hamiltonian density is given by

\be
H = \frac{1}{4\pi} \left(\dot \phi_1^2 + \dot \phi_2^2 + \p_\varphi\phi_1^2 + \p_\varphi \phi_2^2 \right) - \frac{\l}{2\pi} (\dot \phi_1 \dot \phi_2 + \p_\varphi \phi_1 \p_\varphi  \phi_2)
\ee
and the energy evaluates to

\be
E = \frac{\pi}{2R}  \left[(n_1-w_1)^2 +  (n_2+w_2)^2 \right] + \frac{Q_1^2 + Q_2^2}{2\pi R}
\ee
Thus, $E-\frac{Q_1^2 + Q_2^2}{2\pi R}$ is independent of $\l$. This is precisely what we obtained in \eqref{jjsf}, upon taking into account the different normalization of the currents.

The lessons we extract from this small exercise are that while the original $U(1)$ symmetries are still present and the associated charge does not vary along the flow, the current that implements them becomes non-chiral. Restricting to the chiral component of the current makes the charge depend on $\mu/\l$. In defining the chiral charges, one needs to be careful about normalization. It is not hard to see in conformal perturbation theory that $J_1$ acquires an antichiral component proportional to $- \l \bar J_2 $, while the normalization of the original current $J_1$ starts to depend on $\l$ at second order. This indicates in particular that  conformal perturbation theory recreates the fully conserved current $\hat J_1$.

\subsection{Holography}

We would now like to study the same deformation from the point of view of holography. We model the two chiral currents via two Chern-Simons gauge fields of opposite level

\be
S = \frac{k}{8\pi} \int d^3 x \sqrt{g}\,  \e^{\mu\nu\rho} (A_\mu \p_\nu A_\rho - B_\mu \p_\nu B_\rho)
\ee
The variation of the on-shell action is

\be
\d S = - \frac{k}{8\pi} \int d^2 x \sqrt{\g} \, [ (\g^{\mu\nu} + \e^{\mu\nu}) A_\mu \d A_\nu + (\g^{\mu\nu} - \e^{\mu\nu}) B_\mu \d B_\nu] -  \frac{1}{2} \int d^2 x \sqrt{\g} \, (T_{\a\b}^A + T_{\a \b}^B) \, \d \g^{\a\b}
\ee
where we imposed the natural boundary conditions that fix $A_-$ and respectively $B_+$ on the boundary. The stress tensor contribution is

\be
T_{\a\b}^A = \frac{k}{8\pi} (A_\a A_\b - \frac{1}{2} \g_{\a\b} A^2)
\ee
and similarly for $T_{\a\b}^B$. Setting the metric to be flat and $\e^{+-}=2$, we have

\be
\d S = - \frac{k}{4\pi} \int d^2 z (A_+ \d A_- + B_-\d B_+ )
\ee
and the dual currents are given by

\be
J_+ = \frac{k}{4\pi} A_+ \;, \;\;\;\;\; K_- = \frac{k}{4\pi} B_-
\ee
We would now like to add a double trace deformation of the form $-\hat \l \int d^2 z  \, J_+ K_-$ to the on-shell action which, as discussed in section \ref{revdth}, corresponds to adding $+\hat \l \int d^2 z  \, J_+ K_-$ to the CFT action. The variation of the action now reads

\be
\d S = - \frac{k}{4\pi} \int d^2 z \left[A_+ \, \d \left(A_- + \lt B_-\right) + B_- \,\d \left(B_+ + \lt A_+ \right)  \right] \;, \;\;\;\;\;\; \lt= \frac{\hat \l k}{4\pi}
\ee
The relation between $\hat \l$ and $\l$ of the previous section (defined for $k=1$) is $\hat \l = 4 \pi \l$, and so $\tilde \l = \l$. The addition of the double-trace term corresponds to a shift in the sources

\be
A_- \r \tilde{\mathrm{a}}_-= A_- +  \lt  B_- = A_- + \hat \l K_-\;, \;\;\;\;\;\;\; B_+ \r \tilde{\mathrm{b}}_+ = B_+ + \lt  A_+ = B_+ + \hat \l J_+
\ee
while the vevs remain the same. In terms of the new holographic data $\tilde{ \mathrm{a}}_\a, \tilde{\mathrm{b}}_\b$, the asymptotic expansion of the gauge fields is

\be
A = \frac{4\pi}{k} J_+ dx^+ + (\tilde{\mathrm{a}}_- - \hat \l  K_-) dx^- \;, \;\;\;\;\;\; B= \frac{4\pi}{k} K_- dx^- + (\tilde{\mathrm{b}}_+ - \hat \l  J_+) dx^+
\ee
When $\tilde{\mathrm{a}}_- = \tilde{\mathrm{b}}_+ =0$, from the point of view of the original CFT, it looks like there is a non-zero external gauge field, $A_- = - \hat \l K_-$, $B_+ = - \hat \l J_+$. However, from the point of the deformed CFT, it may be more natural to interpret the terms proportional to $\hat \l$ as components of two non-chiral conserved currents, $\hat J = \frac{k}{4\pi} A$ and $\hat K = \frac{k}{4\pi} B$, which have 
%
%

\be
\hat J_+ =J_+;, \;\;\;\;\;\; \hat J_- = - \lt K_- \;, \;\;\;\;\; \hat K_+ = - \lt J_+\;, \;\;\;\;\; \hat K_- =K_-
\ee
%
%
Note this precisely equals \eqref{j1}, under the identification    $J_+ \leftrightarrow \frac{1}{2\pi} \p\phi_1$ and $K_ -\leftrightarrow \frac{1}{2\pi} \bar \p \phi_2$ in the previous subsection. To see this identification even better, one can compute the two-point function of the current using the Chern-Simons equations of motion

\be
\langle J_+ J_+ \rangle_{\mu} = \left. \frac{\d \langle J_+ \rangle}{\d \mathrm{a}_-} \right|_{\mathrm{a}_-=0}\!\!\!\! = \frac{1}{1-\lt^2} \langle J_+ J_+ \rangle_0
\ee
The charge associated to $J_+$ is no longer conserved, though the charge associated to the full $\hat J_\mu$ is, and thus

\be
R \, \frac{k}{4\pi} (A_+ + \lt  B_-) = n_1 \;, \;\;\;\;\; R \, \frac{k}{4\pi} (B_- +  \lt A_+) = -  n_2 
\ee
The charges associated to the normalized chiral currents $ J_+ \sqrt{1-\lt^2}$ and $K_-\sqrt{1-\lt^2} $ (analogues of $\p \tilde\phi_1$ and $ \bar\p \tilde\phi_2$) are

\be
Q_1 = R \, J_+ \, \sqrt{1-\lt^2}  = \frac{n_1 + \lt n_2}{\sqrt{1-\lt^2}} \;, \;\;\;\;\;\;\;\;\; Q_2 =-  R \, K_- \, \sqrt{1-\lt^2}  = \frac{n_2 + \lt n_1}{\sqrt{1-\lt^2}}
\ee
and so it follows that $Q_1^2-Q_2^2$ is constant along the flow. This is the same as the solution \eqref{flowqjj}, provided we identify $\lt = \tanh \mut$.

 Let us also show that the energy agrees with the field theory result. The contribution of the double trace deformation to the stress tensor can be computed from the variational principle, keeping $J^\a$ fixed. The covariant form of the double-trace term is

\be
S_{d.tr} = - \frac{\hat \l}{2} \int d^2 x \sqrt{\g} \, J^\a (\g_{\a\b} + \e_{\a\b}) K^\b =  -\frac{\hat \l k^2}{32\pi^2} \int d^2 x \sqrt{\g} \, A_\a (\g^{\a\b} + \e^{\a\b}) B_\b  
\ee
The variation of this term with respect to the metric keeping  $A_\a,B_\b$ fixed  
 gives the following  contribution to the stress tensor 

\be
\Delta T_{\a\b} =  \frac{\hat \l k^2}{16 \pi^2} \left(A_{(\a} B_{\b)} - \frac{1}{2} \g_{\a\b} \, A_\g B^\g \right)
\ee
The total stress tensor is then

\be
T_{++} = \frac{k}{8\pi} \left(A_+^2 + B_+^2+ \frac{\l k}{2\pi} A_+ B_+ \right) = \frac{k}{8\pi} A_+^2 (1-\tilde \l^2 ) = \frac{2\pi Q_1^2}{k R^2}
\ee
where we used the condition $\mathrm{b}_\a=0$ to equate $B_+ =  - \tilde \l A_+$. Similarly, we have

\be
T_{--} = \frac{k}{8\pi} (B_-^2 + A_-^2 +2 \lt  A_- B_-)= \frac{k}{8\pi} B_-^2 (1-\tilde \l^2 ) = \frac{2\pi Q_2^2}{k R^2}
\ee
Consequently

\be
E_L - \frac{2\pi Q_1^2}{k R} = E_R - \frac{2\pi Q_2^2}{k R} =0
\ee
is constant, as expected. 


	\providecommand{\href}[2]{#2}\begingroup\raggedright\endgroup
	

\end{document}